\documentclass[aps,preprint,superscriptaddress,raggedbottom,longbibliography]{revtex4-1}
\usepackage{graphicx}
\usepackage{amsmath}
\usepackage{amssymb}
\usepackage{color}
\usepackage{amscd}
\usepackage{enumerate}

\usepackage{gensymb}
\usepackage{mathrsfs}
\usepackage{calrsfs}
\usepackage{epsfig}
\usepackage{subfigure}
\usepackage{psfrag}
\usepackage{amssymb}

\begin{document}

\title{A cycling state that can lead to glassy dynamics in intracellular transport}
\author{Monika Scholz}
\affiliation{James Franck Institute, the University of Chicago, Chicago, IL 60637}
\affiliation{Institute for Biophysical Dynamics, the University of Chicago, Chicago, IL 60637}
\author{Stanislav Burov}
\affiliation{Department of Physics, Bar-Ilan University, Ramat-Gan, 5290002 Israel}
\author{Kimberly L.\ Weirich}
\affiliation{James Franck Institute, the University of Chicago, Chicago, IL 60637}
\affiliation{Institute for Biophysical Dynamics, the University of Chicago, Chicago, IL 60637}
\author{Bj\"orn J. Scholz}
\affiliation{Enrico Fermi Institute, the University of Chicago, Chicago, IL 60637}
\affiliation{Kavli Institute for Cosmological Physics, the University of Chicago, Chicago, IL 60637}
\affiliation{Department of Physics, the University of Chicago, Chicago, IL 60637}
\author{S.\ M.\ Ali Tabei}
\affiliation{Physics Department, University of Northern Iowa Cedar Falls, Iowa 50614}
\author{Margaret L.\ Gardel}
\affiliation{James Franck Institute, the University of Chicago, Chicago, IL 60637}
\affiliation{Institute for Biophysical Dynamics, the University of Chicago, Chicago, IL 60637}
\affiliation{Department of Physics, the University of Chicago, Chicago, IL 60637}
\author{Aaron R.\ Dinner}
\thanks{To whom correspondence may be addressed:  dinner@uchicago.edu}
\affiliation{James Franck Institute, the University of Chicago, Chicago, IL 60637}
\affiliation{Institute for Biophysical Dynamics, the University of Chicago, Chicago, IL 60637}
\affiliation{Department of Chemistry, the University of Chicago, Chicago, IL 60637}

\begin{abstract}
Power-law dwell times have been observed for molecular motors in living
cells, but the origins of these trapped states are not known. We
introduce a minimal model of motors moving on a two-dimensional network
of filaments, and simulations of its dynamics exhibit statistics
comparable to those observed experimentally. Analysis of the model
trajectories, as well as experimental particle tracking data, reveals a
state in which motors cycle unproductively at junctions of three or more
filaments. We formulate a master equation for these junction dynamics
and show that the time required to escape from this vortex-like state
can account for the power-law dwell times. We identify trends in the
dynamics with the motor valency for further experimental validation. We
demonstrate that these trends exist in individual trajectories of myosin
II on an actin network. We discuss how cells could regulate
intracellular transport and, in turn, biological function, by
controlling their cytoskeletal network structures locally.

\end{abstract}

\thanks{To whom correspondence may be addressed:  dinner@uchicago.edu}

\maketitle

\section{Introduction}
Individual microscopic particles (beads \cite{wong2004anomalous, wang2009anomalous} or fluorescently labeled molecules \cite{parry2014bacterial,sako2000single,cai2009single}) can now be tracked in cells.  These studies reveal complex dynamics \cite{zimmerman1991estimation,konopka2006crowding,hofling2013anomalous}.  The resulting trajectories can be treated as random walks, and quantitative analysis of their statistics can provide insights into underlying mechanisms \cite{tabei2013intracellular,burov2013distribution}. Often, the mean square displacement (MSD) exhibits a power-law (typically sublinear) dependence on the separation in time between two observations, known as the lag time ($\Delta$) \cite{CoxPRL, tolic2004anomalous, weber2010bacterial, weigel2011ergodic, tabei2013intracellular,burov2013distribution,bressloff2013stochastic}. 
In certain cases \cite{tabei2013intracellular,weigel2011ergodic}, the MSD also decays as the amount of data included in averages (the measurement time, $T$) increases; this trend indicates a power-law distribution of dwell times and is known as ``aging'' in theories of glassy dynamic \cite{cugliandolo1994evidence}.

These correlations can have important biological implications \cite{tabei2013intracellular,manzo2015weak}.  For example, a recent study shows that the anomalous dynamics observed for insulin secretory vesicles (granules) can account for the biphasic kinetics of insulin release \cite{tabei2013intracellular} without the need to invoke separate pools of granules, as previously \cite{Seino2011}.  In particular, the sustained release relies on the glassy dynamics.  Glassy dynamics are often interpreted in terms of trapping in local minima of an energy landscape with an exponential or power-law distribution of depths \cite{bouchaud1995aging, feigel1988stochastic}.  However, how such a landscape could arise from typical biomolecular interactions is unclear.   Crowding is insufficient, as it results in standard Brownian motion but with a reduced diffusion coefficient \cite{Dix2008}.  Because the moving vesicles are associated with molecular motors, which consume cellular energy stores (nucleotide triphosphates) for directed motion along cytoskeletal filaments, other, intrinsically nonequilibrium mechanisms of generating these statistics may exist.

\begin{figure}[h!]
\begin{center}
\includegraphics[width=0.75\textwidth]{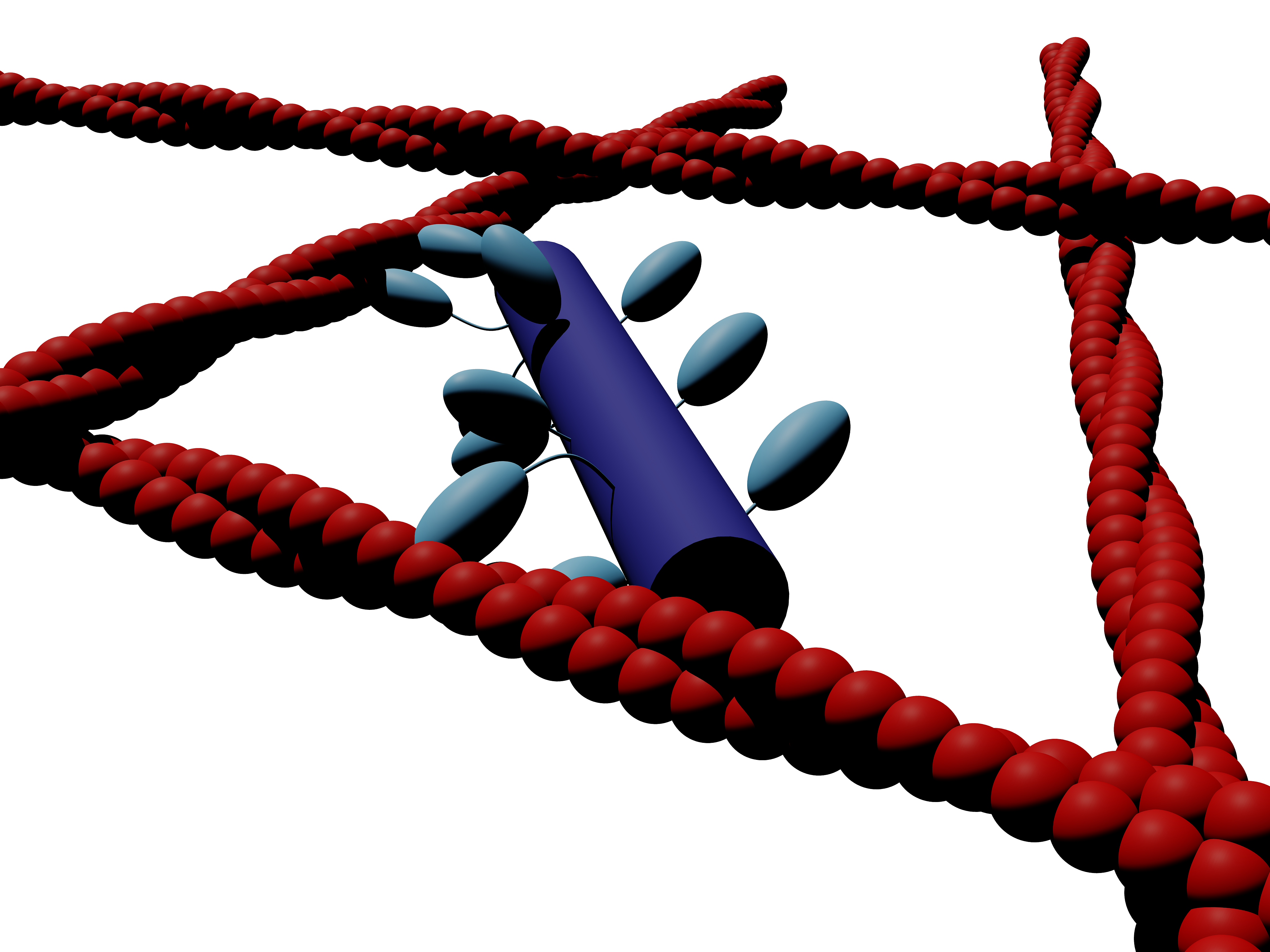}
\caption{Schematic of a cytoskeletal assembly comprised of multiple molecular motors.  Drawing is to scale for actin filaments (red) and myosin motors (blue).
 \label{fig:actin}%
}
\end{center}
\end{figure}

In the case of insulin release, the vesicles have both kinesin and dynein associated with them \cite{varadi2003kinesin}, which walk in opposite directions on microtubules \cite{vale1992directional}.  More generally, many cytoskeletal assemblies in cells have multiple motors associated with them (Fig.\ \ref{fig:actin}).  Thus it is natural to ask if glassy dynamics could arise from a tug-of-war.  Mathematical models of competing motors have been formulated and show that, depending on the number of motors, their properties, and their binding affinities, different regimes of transport kinetics can be accessed \cite{muller2008tug,newby2010local}.
While the tug-of-war model can explain bidirectional transport in a diverse set of biological systems \cite{hendricks2010motor, soppina2009tug, ali2011myosin, hancock2014bidirectional}, standard formulations cannot account for aging because  exponential dwell times are assumed. 
Moreover, the tug-of-war is essentially a single-filament mechanism.  Cytoskeletal networks contain geometric structures that involve multiple filaments (e.g., junctions), and these could support other dynamics.

In this paper, we investigate the dynamics of a minimal model of a motor that can make multiple attachments to a two-dimensional network of filaments.  Using simulations, we show  that such a model can exhibit glassy dynamics, and we discover that the long-time correlations in this model result from vortex-like trajectories that motors follow when three or more filaments cross to form a circuit.  This represents a new mechanism for trapping that does not require individual motor heads stalling, and we term it the ``cycling state''.  We obtain average flows for idealized junction geometries from a master equation analysis and show that trapping in vortex-like cycling states can give rise to glassy, non-ergodic dynamics like those observed in experiments.  We analyze experimental particle tracking data to demonstrate the presence of the cycling state in measured trajectories and show that it relates to exponents that quantify aging.  Broader implications for biological function are discussed.

\begin{figure}[h!]
\begin{center}
\includegraphics[width=0.75\textwidth]{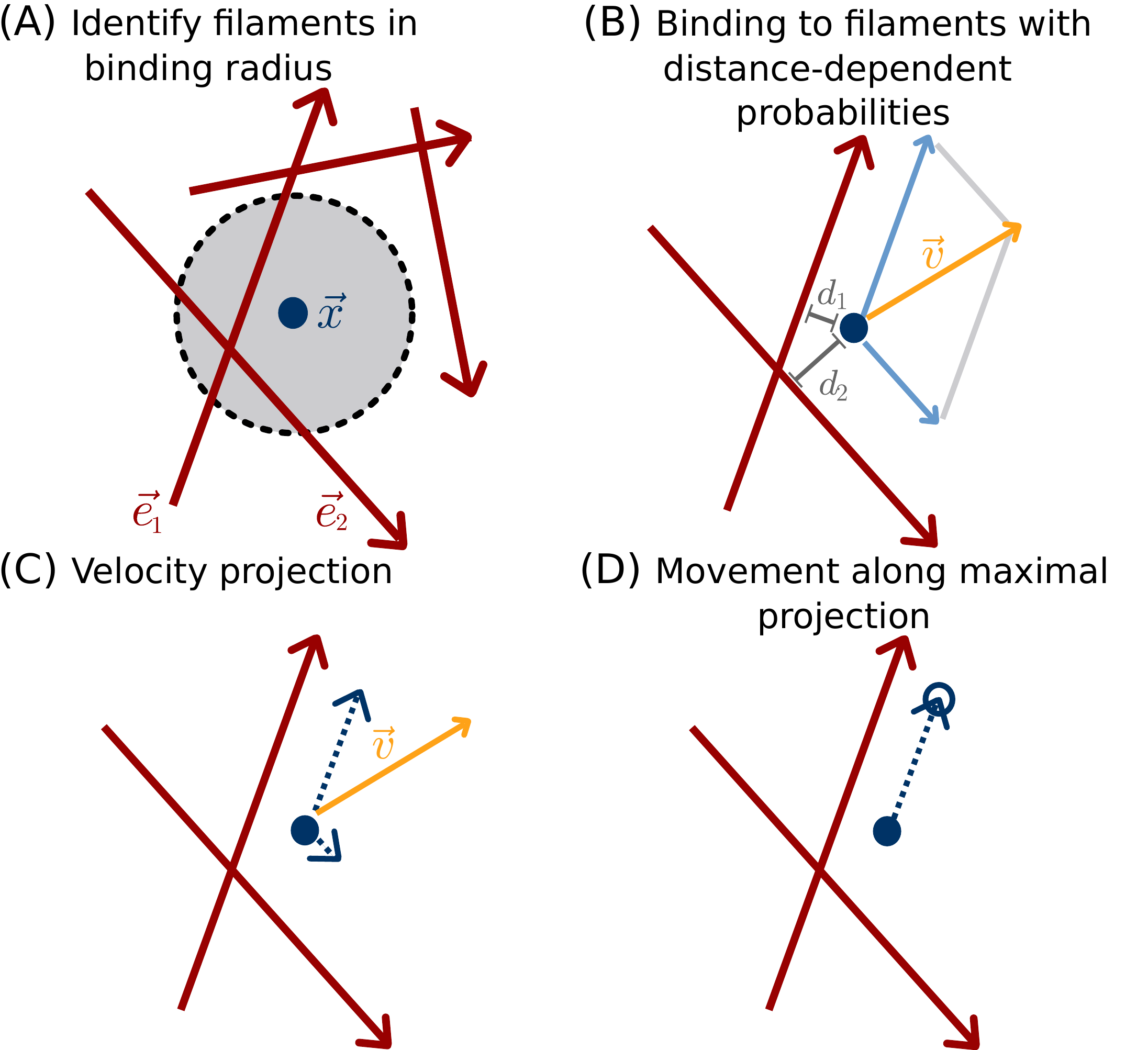}
\caption{Schematic of the molecular simulation procedure.
(A) The motor (blue dot) can associate with filaments (red vectors) within a defined binding radius.  (B) Active attachments are assigned stochastically in a distance-dependent fashion.  The velocity vector that results from each active attachment moving a step along its filament is computed.  (C) The velocity vector is projected onto the filaments.  (D) The motor takes a step with size proportional to the magnitude of the largest velocity projection, along the associated filament.
\label{fig:schematic}
}
\end{center}
\end{figure}

\section{Molecular simulations}

We model filaments as randomly oriented line segments in a plane.  The length of each filament is drawn from a normal distribution with a maximum length $\ell$ and standard deviation $\sigma_{\ell}$.  We associate the polarity of filament $i$ with a fixed unit vector $\vec{e_i}$.  The filaments are static, and thus represent experimental situations in which cytoskeletal rearrangements are slow in comparison with the period over which motor transport is measured. 
For simplicity, we also neglect heterogeneities in the composition of the filaments and the solution environment, which can lead to complex dynamics \cite{newby2010local}.

We are interested in cases in which many molecular motors act in concert---e.g., a vesicle with several protein motors attached or a myosin minifilament.  
We refer to our model of such an assembly as ``a motor''.  A motor is a point particle that can bind up to $M$ filaments at once and move along them as follows. We separate the binding process into two steps (Fig.\ \ref{fig:schematic}, panels 1 and 2; see SI Text and Fig.\ S1 for an alternative scheme). First, we distribute the $M$ possible attachments for a motor among filaments with probability proportional to $b_i = \exp[-(3 d_i/2 s)^2]$, where $d_i$ is the shortest distance between the motor and filament $i$, and $s$ is a parameter that sets the interaction length scale.  This probability is normalized by the sum $\sum_ib_i$ over all filaments within a distance $3s$ of the motor.  Then, we determine if each such interaction exerts force to move the motor (henceforth, ``active'') with probability $b_i$ or not (``inactive'') with probability $1-b_i$.   We denote the number of active attachments to filament $i$ by $k_i$.  To determine the change in position of the motor we compute the vector $\vec{v} = \sum_i k_i  \vec{e_i}$ (orange  in Fig.\ \ref{fig:schematic}B).  This choice is consistent with measurements that show that motor velocities increase with head numbers \cite{lee2015axonal}.  We project $\vec{v}$ onto all the filaments with at least one active attachment and add to the motor position the projection with the maximum magnitude scaled by the time step (Fig.\ \ref{fig:schematic}C and D). This projection rule ensures that the motor moves along rather than off filaments.   It also gives rise to an effective force-velocity dependence, with opposing parallel velocities canceling each other.   We assume that forces that are directed orthogonally do not create a load on the motor. 
For the simple scenario of orthogonal filaments, this scheme thus simplifies to a step in the direction of the filament with the highest number of active binding interactions (i.e., a majority rule). The projection rule implies that binding sites do not detach under load; rather, they stay bound and contribute to the overall velocity vector. Simulations relaxing the projection rule to a simple net velocity calculation show only minor differences (Fig.\ S2).   Simulations with stochastic selection between projections with probabilities proportional to their magnitudes also yielded similar results.

\begin{table}
[tb]
%\begin{ruledtabular}
\begin{tabular}{lr} \hline
Parameter& Value \\ \hline
Filament density per unit area & 1\\
Filament length $\ell$&5\\
Filament standard deviation $\sigma_\ell$& 5\\
Binding radius $s$ & 0.01\\
Total number of binding sites $M$ &50-100\\
Time step $dt$ & 0.001\\
Total number of steps $T$ & $10^5$\\
Trials & 2000\\ \hline
\end{tabular}
%\end{ruledtabular}
\caption{Simulation parameters}
\label{tab:sim_parameters}

\end{table}

\begin{figure}[h!]
\begin{center}
\includegraphics[width=0.75\textwidth]{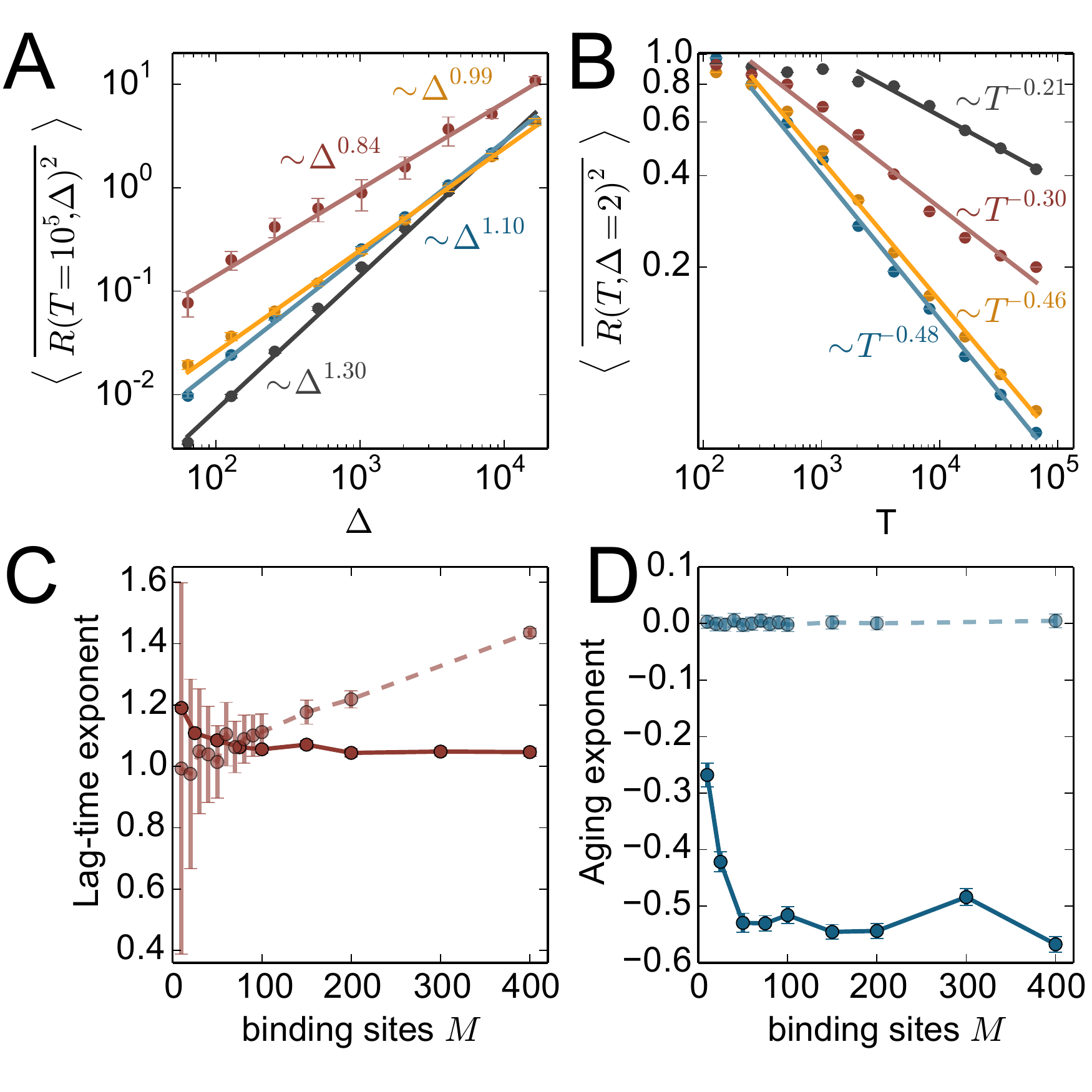}
\caption{
Scaling in molecular simulations.  (A) Time-averaged MSD as a function of lag time for binding radii $s = 0.001$ (gray), 0.01 (blue), 0.1 (orange) and 1(red). 
(B) Mean-squared displacement as a function of measurement time for a range of binding radii (colors are the same as in (A)).  Each curve was rescaled to start at 1 for easier visualization.
(C) and (D) Dependence of the indicated exponents on the maximum number of possible attachments, $M$. The dashed lines show the outcome for a tug-of-war scenario, which can be obtained from our model by simulating the dynamics on two antiparallel filaments. The full lines show the exponents for the molecular simulations on a two-dimensional random network. Lag-time exponents were calculated for $T = 10^5$.
\label{fig:msd}
}
\end{center}
\end{figure}

The values of the simulation parameters are given in Table \ref{tab:sim_parameters}.  While the model is general, we choose the values to be roughly consistent with actin and myosin to ensure that we study a physically reasonable regime.  To this end, we assume a myosin speed of 1 $\mu$m/s, which is in the range of speeds reported from \textit{in vitro} and \textit{in vivo} studies of various classes of myosin \cite{finer1994single, howard1997molecular,Pierobon20094268}. We associate our unit length with 1 $\mu$m, such that the average filament length is 5 $\mu$m. The binding radius of the motor is then about 10 nm, and a single time step of the simulation is 0.1 s. Although the properties of actual molecular motors vary substantially, our conclusions are robust to parameter choices that range over an order of magnitude (Fig.\ \ref{fig:msd}).

We simulate the model according to the rules above and calculate the time-averaged mean square displacement (MSD) for the resulting trajectories:

\begin{equation}
\langle\overline{\vec{R}(T,\Delta)^2}\rangle=\frac{1}{T-\Delta} \int_0^{T-\Delta} [\vec{x}(t+\Delta)-\vec{x}(t)]^2 dt,
\label{eq:1}
\end{equation}
where $\vec{x}(t)$ is the position of a motor at time $t$.  We  plot the time-averaged MSD as a function of lag time ($\Delta$)  in Fig.\ \ref{fig:msd}A.
We use the time-averaged MSD  to make connection with biological experiments that typically have a limited number of trajectories \cite{saxton1997single}.  It is important to note that the time- and ensemble-averaged MSDs can exhibit different scalings when the process of interest is non-ergodic \cite{bouchaud1992weak, lubelski2008nonergodicity, he2008random, sokolov2008viewpoint, schulz2013aging}.

 By varying the binding radius $s$ by factors of 10 from 0.001 to 1, we can tune the transport from superdiffusive to subdiffusive.  When the range of interaction is very small, the motor only attaches to the closest filament.  As a result, the motion is ballistic but slow (note the intercept for the gray line in Fig.\ \ref{fig:msd}A) because the number of active attachments is low.  For intermediate interaction ranges, the motor can simultaneously bind multiple filaments, and a tug-of-war-like mechanism gives rise to subdiffusive or diffusive motion.

We plot the MSD as a function of the measurement time ($T$) in Fig.\ \ref{fig:msd}B.  For an ergodic system, the MSD should be constant as $T$ varies---the properties of the motion are the same independent of the length of recording. In contrast, a decrease in the MSD with $T$ suggests that there are long-lived traps.  The essential idea is that the traps increasingly dominate the statistics as more data are included in the averages.  We observe a power-law decay (i.e., aging), consistent with prior analysis of myosin II on an \textit{in vitro} actin network \cite{burov2013distribution} and the motion of insulin granules \textit{in vivo} \cite{tabei2013intracellular, burov2013distribution}.

The observed aging exponent depends on the size of the binding radius $s$, but we obtain exponents comparable to experimental values as this parameter varies over two orders of magnitude. The molecular simulations of the model also predict more trapping for increased numbers of binding sites: both the lag-time and the aging exponents decrease with larger $M$ (Fig.\ \ref{fig:msd}C and D). 
However, due to the non-ergodic nature of this transport process, the lag-time exponent depends on the length of the recording $T$. We show the $T$-dependence of the exponents in Fig.\ S3 (see also Fig.\ S4).  
In contrast to the results that we obtain for motion on a random filament network, a simple tug-of-war scenario with a motor between two antiparallel filaments yields no glassy dynamics, independent of the number of binding sites. Also, the scaling of the MSD becomes increasingly ballistic with increased numbers of binding sites (Fig.\ \ref{fig:msd}C and D), and the diffusion constant increases (data not shown).

\begin{figure}[h!]
\begin{center}
\includegraphics[width=0.75\textwidth]{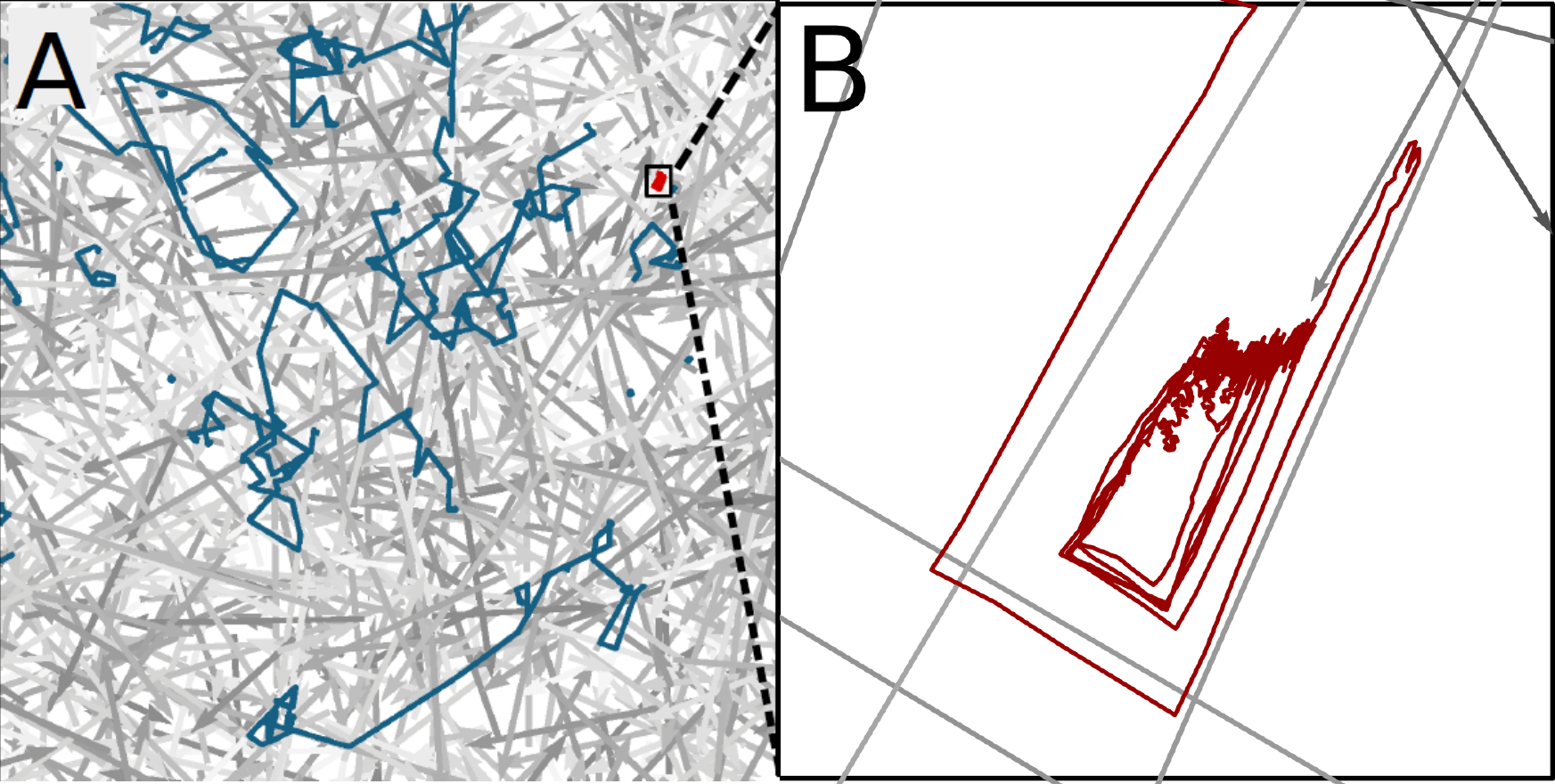}
\caption{Single-particle trajectories reveal a cycling state. (A) Representative simulation trajectories (blue and red) projected onto the filament network (gray).  (B) A magnified view of a cycling trajectory (red). 
\label{fig:trajectories}
}
\end{center}
\end{figure}

Having thus captured the statistics of experiments, we sought to use the model to elucidate the microscopic motions that underlie the statistics.  In this regard, we noticed that motors frequently exhibit vortex-like motions in which they steadily cycle from one filament to another at a junction (Fig.\ \ref{fig:trajectories}A).  These motions persist for long times in comparison with the duration of the simulations.  Cycling motions are of particular interest given that passive particles in a vortex flow in a fluid are known to exhibit power-law-distributed trapping times \cite{solomon1993observation}.  Analogous observations exist for trapped ions \cite{devoe2009power} and Bose-Einstein-Condensates \cite{lundh2000hydrodynamic}.

An approximate length scale for the vortices responsible for the glassy dynamics can be inferred from the crossover in the MSD curves. When the MSD as a function of lag time switches from super-linear to linear/sublinear scaling, $\Delta$ is sufficiently large for the MSD to include contributions from the vortices (Figs.\ S1 and S2). In other words, the MSD exponent decreases when the dynamics include bounded trajectories. In our conditions, the crossover occurs at $\Delta = 100$, which corresponds to about 10 s based on the numbers for actin and myosin given above. In turn, for a myosin II motor and the parameters in Table \ref{tab:sim_parameters}, the length scale of vortices would be approximately 100 nm.

\section{Cycling state gives rise to power-law dwell times}

\begin{figure}[h!]
\begin{center}
\includegraphics[width=0.75\textwidth]{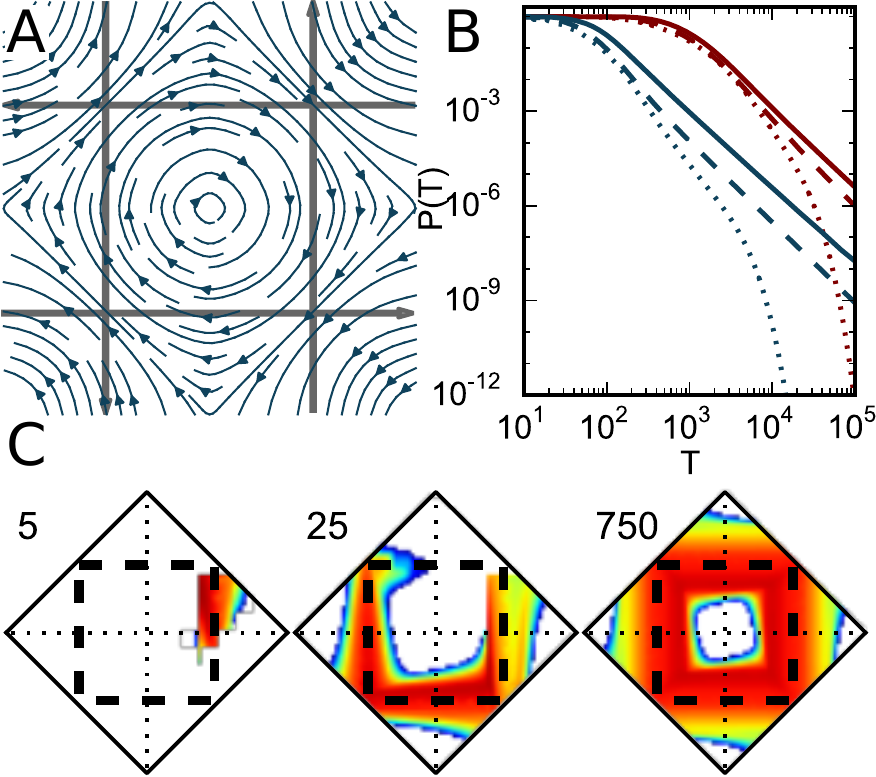}
\caption{
Vortex model.  (A) Streamlines (blue) for motion in the vicinity of an idealized square circuit of filaments (gray).  (B) Survival probability in a vortex for $M=1$ (red) or $M=5$ (blue) in a vortex of size  40. The probability is initialized  3/4 of the diagonal.  The binding radius changes from 1/5, 1/7.5 or 1/10 of the vortex length. (dotted, dashed or full lines).
The probability decays in three phases: constant, power-law, and exponential. The length of the power-law decay decreases with increasing binding radius. (C) Numerical solution of the master equation in a square vortex. In the example shown, the initial condition is a localized probability at a single location similar to B.
The vortex size is 40, with $M$=5, a binding radius of 1/10 of the side length and a unit length step size. The colors correspond to the logarithm of the probability density of observing the motor at a specific location (red is highest) and is normalized in each subplot. The three panels show the master equation after 5, 25, and 750 iterations, respectively. 
 The probability density spreads over time and the maximum rotates. The final image shows a pattern of localization that is stable as the integrated density continues to decrease.  
\label{fig:ploss}%
}
\end{center}
\end{figure}

To investigate whether the observed vortex-like motion can account for the anomalous statistics, we now consider an idealized geometry.   Specifically, we consider filaments that meet at right angles to form a square because it simplifies the mathematics (Fig.\ \ref{fig:ploss}). However, we emphasize that the conclusions that we draw from this analysis are general---cycling can occur whenever the unit vectors of a group of crossing filaments sum to zero. In a square loop, the motor only interacts with one or two filaments at a time, and, due to the projection rule (Fig.\ \ref{fig:schematic}), the motor always moves along the filament with the majority of sites bound. The average resulting motion can be described by streamlines, which form a nested set of closed loops that do not cross (Fig.\ \ref{fig:ploss}A).  If the dynamics were deterministic,  the streamlines would describe the motion entirely, and the motor would stay in the vortex forever. However, due to the stochastic nature of binding and unbinding, individual trajectories deviate from the streamlines, and the motor eventually leaves the vortex.

To characterize this behavior quantitatively, we derive the  master equation that governs this escape.  For this purpose, we need to determine how the active binding sites distribute between the two accessible filaments via the two-step procedure described in {\it Molecular Simulations}. The probability $P_f$ that $l_1$ out of $M$ possible attachment sites are assigned to filament 1 and that the remaining are assigned to filament 2 is binomial:

\begin{equation}
P_f(l_1,l_2)= \binom{M}{l_1} \frac{b_1^{l_1} b_2^{l_2}}{(b_1+b_2)^M} , \textrm{with}\ l_1+l_2 =M
\end{equation}
The probability $P_{a}$ that $k_1$ out of $l_1$ possible attachments are active is also binomial:

\begin{equation}
P_{a}(k_1|l_1)= \binom{l_1}{k_1} b_1^{k_1} (1-b_1)^{l_1-k_1},
\end{equation}
and similarly for $k_2$.
Since filament assignment and selection of the active attachments are independent events, we can now write for the overall probability $P$ of having $k_1$ and $k_2$ active binding sites on filaments 1 and  2

\begin{equation}
P(k_1,k_2)=\sum_{l_1=k_1}^{M-k_2} P_f(l_1,l_2) P_{a}(k_1|l_1)P_{a}(k_2|l_2).
\label{eqn:P}
\end{equation}
The limits of the sum are set by $l_1\geq k_1$, $l_2\geq k_2$, and $l_1+l_2 = M$.

Associating filament 1 with the $x$ direction and filament 2 with the $y$ direction, the resulting master equation for motion in a square vortex is

\begin{eqnarray}
\frac{\, dP_t(x,y)}{dt}& =& - \left(1 - \displaystyle\sum\limits_{k_1=k_2}^{M/2}  P(k_1,k_2) \right) P_{t-1}(x,y)\\\nonumber 
&&+ \sum\limits_{k_1 > k_2}^{M - k_2} P(k_1,k_2) P_{t-1}(x-k_1,y) \\ \nonumber 
&&+ \sum\limits_{k_2 > k_1}^{M - k_1} P(k_1,k_2) P_{t-1}(x,y-k_2).
\end{eqnarray}
Here, $P_t(x,y)$ is the probability to be in a certain location $(x,y)$ in the vortex at time $t$, and we assume a unit time step. 
The first term accounts for motors that stay in place, while the second and third terms account for taking steps along filaments 1 and 2, respectively.  Note that the polarity of the filaments is fixed, so that motion along each filament is unidirectional and no terms are needed to account for motion in the other direction.  The vortex has absorbing boundaries where the velocity has a saddlepoint, forming a diamond shaped region.

We solve this master equation numerically (Fig.\ \ref{fig:ploss}C).  The survival probability in the vortex decays in three stages. At short times, the probability is close to unity, since the probability distribution needs a finite number of steps to reach the vortex boundaries. Then, we observe that the probability density also rotates within the vortex, similar to particles in the explicit simulations (see also Supplementary Movie 1). During that time, the probability decays as a power law (see Fig.\ \ref{fig:ploss}B). The probability distribution in the vortex ultimately reaches a quasi-steady-state, when the shape of the distribution is no longer changing, and the survival probability decays as an exponential. The durations of these three stages are determined by the starting conditions and the relative size of the binding radius to the size of the vortex, as well as the number of binding sites. The power-law decay of the survival probability is longest for small binding radii (significantly smaller than the vortex) since the binding potential favors steps parallel to the filaments, and only allows very small steps orthogonal to the filaments.  This relates to the results shown in Fig.\  \ref{fig:msd}A and B, in that there exist a range of vortex sizes in our chosen network, which lead to characteristic trapping times. Binding radii much smaller or larger than these sizes will show less aging than midrange radii, due to the existing spatial scales in the random network.
Relatedly, increasing the density of filaments in the network decreases the vortex size; this leaves the scaling with $\Delta$ unchanged while decreasing the extent of aging (Fig.\ S5).

The average motion in the vortex can be viewed as a combination of drift along the streamlines and diffusion orthogonal to the streamlines. From this perspective, the cycling motion is similar to that of a particle in a Rayleigh-Benard convection cell \cite{shraiman1987diffusive}. However, the situation differs, in that the diffusion is position-dependent (i.e., $P(k_1,k_2)$ varies in space through $b_1$ and $b_2$).  In other words, the master equation limits to a drift-diffusion (Langevin) form with a multiplicative, rather than an additive, noise.  This form reflects the fact that, in the motor model, the number of attachments is influenced by the position relative to the filaments defining the vortex boundary.

\section{Experimental demonstration of cycling state contributions}

A key prediction of the cycling state model is that the aging exponent decreases with the number of attachment sites on each motor (Fig.\ \ref{fig:msd}D).  We can test this idea without the confounding effects of cell signaling by studying mixtures of purified actin filaments and myosin motors. The specific system that we consider comprises actin filaments bundled by a passive crosslinker fimbrin. The motors are minifilaments of skeletal muscle myosin II, which polymerizes into large assemblies with on the order of 100 motor proteins \cite{thoresen2013thick}.  The actin and myosin molecules are visualized through fluorescence microscopy, as detailed in the SI Text. 
Single-particle trajectories were obtained by tracking, and trajectories with fewer than 30 timepoints were discarded.  There are 246 such trajectories with a mean length of 166 s, with a standard deviation of 120 s.  This length is long compared with typical {\it in vitro} particle-tracking studies of isolated motors on single filaments. Both the large number of heads and the density of binding sites in the filament network make it unlikely for motors to detach, which favors processivity.
The mean-squared displacement as a function of measurement time $T$ shows aging, with an exponent comparable to previously published data (compare \cite{burov2013distribution} with Figs.\ S6A and B).  

\begin{figure}[h!]
\begin{center}
\includegraphics[width=0.75\textwidth]{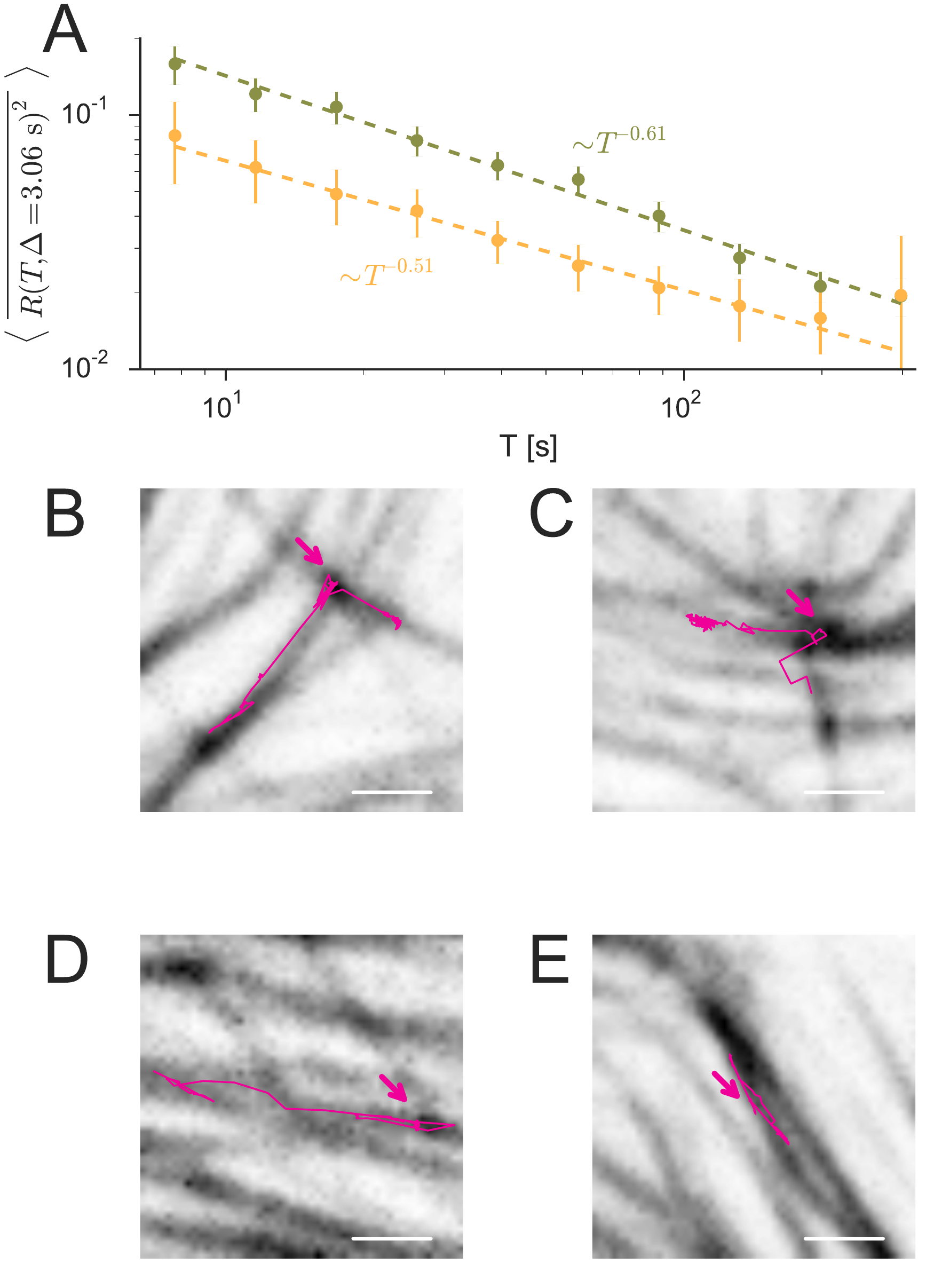}
\caption{Experimental trajectories show cycling. (A) Trajectories were divided into two groups according to the total fluorescence intensity of the motor. The group with higher fluorescence intensity (green) shows a larger aging exponent than the low-fluorescence group (yellow). 
(B-E) Representative trajectories of individual myosin II minifilaments overlayed onto the actin network. Arrows denote cycling events. Trajectories were obtained from myosin II minifilaments on an actin network bundled with fimbrin.  Scale bar is 1.1 $\mu$m; trajectories shown are imaged at 1.5 s intervals; see SI Text for details. Single-particle trajectories were obtained using the Python-based implementation of the Crocker-Grier algorithm Trackpy \cite{daniel_b_allan_2014_9971}. \label{fig:fimbrin} 
}
\end{center}
\end{figure}

To test the prediction in Fig.\ \ref{fig:msd}D, we exploit the fact that the number of myosin
molecules in each minifilament varies naturally and use fluorescence intensity as a proxy for the number of motor heads. We divide the trajectories into two groups according to the median fluorescence of the motor.
There are 132 single-particle trajectories for motors with relatively high intensity and an equal number of trajectories for
motors with relatively low intensity.  We expect motors with higher intensity to have more heads and thus exhibit stronger aging.  We observe that this is the case (Fig.\ \ref{fig:fimbrin}A).  
Specifically, we used case resampling (a form of bootstrapping) to get a distribution of exponents from each fluorescence group. The means of these distributions were $-0.6125 \pm 0.0035$ and $-0.5112 \pm 0.0086$ (mean $\pm$ SEM high fluorescence and low fluorescence groups, respectively).  We tested if the means are significantly different using the unequal variances $t$-test. The two-tailed $p$-value is less than  10$^{-4}$, indicating that the difference is significant.   

Additionally, we manually selected trajectories that visibly cycled (looped over the same multipixel region more than once).
29 trajectories were found to cycle, and a significant portion of these were also classified as high-fluorescence ($N = 22$, $p$-value = 0.00185, Fisher's exact test for cycling/non-cycling vs.\ high/low fluorescence). Because the actin filaments do not move significantly (Fig.\ S6), we can project trajectories onto the network, and we see that the trajectories that result in the most pronounced decay have visible cycling that coincides with filament junctions (arrows in Figs.\ \ref{fig:fimbrin}B-D).  

While the observations above demonstrate the existence of cycling, we sought to exclude alternative mechanisms for the statistical differences in exponents in Fig.\ \ref{fig:fimbrin}A.  Reasonable candidates are intrinsic differences in motor speeds and detachment.  With regard to the former, we note that the average motor step size from frame to frame (i.e., the net speed for movement over the filament network) does not differ for the high and low intensity groups:  the means $\pm$ SEM are $44.1 \pm 2.6$ nm/s  and $45.0 \pm 2.5 $ nm/s, respectively.   If  detachment were instead responsible for the anomalous dynamics, the particles should undergo simple diffusion when apparently trapped.  To test for this possibility, we divided the trajectories between trapped and non-trapped periods and then analyzed their dynamics using a published measure for detecting complex dynamics at resolutions of only a few frames.   We find that the motion is not simple diffusion, as detailed in the SI Text (see also Fig.\ S9).

Together, these results strongly support the idea that the cycling state
exists for motors with multiple attachments and gives rise to power-law decay in the MSD as a function of measurement time.

\section{Discussion}

Motor-driven processes in cells often exhibit anomalous statistics \cite{weiss2004anomalous,amblard1996subdiffusion,tolic2004anomalous, jeon2011vivo}, and these statistics can have important functional consequences \cite{tabei2013intracellular}. 
While existing random-walk models of aging typically start by assuming trapped states with power-law dwell times, we
introduce a microscopic mechanism for how simple biologically consistent interactions can
give rise to such long-lived traps.
The glassy dynamics result from a vortex-like state that emerges in multiple dimensions---motors that can attach to multiple filaments simultaneously cycle unproductively at filament junctions, and the escape times from these flows are power-law-distributed over a time range set by the motor step size and the filament spacing.  We demonstrate that such cycling events occur frequently in the motion of skeletal myosin II assemblies on a dense, random network of bundles of actin filaments {\it in vitro}, and we use this system to validate the predictions of the model.
To the best of our knowledge, these cycling dynamics have not been appreciated previously, and we expect that their topological features will lead to rich physics beyond idealized trap models. 

Microtubule structures with many intersections are observed above the basal membrane of epithelial cells, where they function in endocytic vesicle transport \cite{reilein2005self}. A study that reconstructed these epithelial microtubule networks \textit{in vitro} observed a kinesin-coated bead cycling through a vortex structure \cite{ross2008kinesin}. The same study (and others \cite{osunbayo2015cargo}) also saw a slow-down or pausing states at intersections, which is also present in our model due to the motors interacting with both filaments at the same time. While these structures are macroscopic compared to the vortices responsible for the observed glassy behavior, they show that motor associated cargo or multi-binding complexes can navigate intersections and cycles for certain geometries.

Tug-of-war models \cite{muller2008tug} contain similar molecular elements and exhibit pauses when motors are stalling or opposing forces match exactly.  However, such models are essentially one-dimensional and do not give rise to power-law-distributed dwell times.  Based on Figs.\  \ref{fig:msd}C and D, we expect the cycling state to be prevalent only when motor assemblies comprise many protein motors and directed motion dominates over thermal processes.  Indeed, measured exponents for the MSD could be used to infer the size of such assemblies.  However, care is needed because different filament structures favor different amounts of directed, tug-of-war, and vortex-like motions.  Random networks in three dimensions have few circuits that lead to cycling, while cytoskeletal networks that are organized by specific filament binding proteins (e.g., Arp2/3), as well as quasi-two-dimensional networks that arise in cell cortices, have larger numbers of suitable junctions.  Thus accurate estimates of the numbers of active motor heads requires constructing a calibration curve for the exponent for each network structure.

Understanding how motor assemblies behave on complex filament networks in cells is an outstanding challenge \cite{elting2012future}.
The degree to which the cycling state contributes to dynamics in different contexts {\it in vivo} is an open question deserving further study.
Cells could potentially spatiotemporally control intracellular transport by rearranging their cytoskeletal networks to favor or disfavor cycling.  Understanding how this control mechanism manifests in different types of cells and tissue environments,  as well as its interplay with other regulatory processes and trapping mechanisms \cite{newby2010local}, is a useful direction for future research.  It will also be interesting to understand the interplay of motor transport, force transmission, and network rearrangement.

\section{Acknowledgments}

We thank Samantha Stam and David Kovar's laboratory for protein purification; Michael Murrell, Jennifer Ross, Toan Huynh, and Norbert Scherer for helpful conversations; and Glen Hocky and Shiladitya Banerjee for critical readings of the manuscript. 
Support was provided by the University of Chicago Materials Research Science and Engineering Center (NSF DMR-1420709) and the W. M. Keck Foundation.

\clearpage

%\bibliographystyle{apsrev4-1}
%\bibliography{converted_to_latex.bib}

\begin{thebibliography}{10}%
\makeatletter
\providecommand \@ifxundefined [1]{%
 \ifx #1\undefined \expandafter \@firstoftwo
 \else \expandafter \@secondoftwo
\fi
}%
\providecommand \@ifnum [1]{%
 \ifnum #1\expandafter \@firstoftwo
 \else \expandafter \@secondoftwo
\fi
}%
\providecommand \enquote [1]{``#1''}%
\providecommand \bibnamefont  [1]{#1}%
\providecommand \bibfnamefont [1]{#1}%
\providecommand \citenamefont [1]{#1}%
\providecommand\href[0]{\@sanitize\@href}%
\providecommand\@href[1]{\endgroup\@@startlink{#1}\endgroup\@@href}%
\providecommand\@@href[1]{#1\@@endlink}%
\providecommand \@sanitize [0]{\begingroup\catcode`\&12\catcode`\#12\relax}%
\@ifxundefined \pdfoutput {\@firstoftwo}{%
 \@ifnum{\z@=\pdfoutput}{\@firstoftwo}{\@secondoftwo}%
}{%
 \providecommand\@@startlink[1]{\leavevmode}%
 \providecommand\@@endlink[0]{}%
}{%
 \providecommand\@@startlink[1]{%
  \leavevmode
  \pdfstartlink
   attr{/Border[0 0 1 ]/H/I/C[0 1 1]}%
   user{/Subtype/Link/A<</Type/Action/S/URI/URI(#1)>>}%
  \relax
 }%
 \providecommand\@@endlink[0]{\pdfendlink}%
}%
\providecommand \url  [0]{\begingroup\@sanitize \@url }%
\providecommand \@url [1]{\endgroup\@href {#1}{\urlprefix}}%
\providecommand \urlprefix [0]{URL }%
\providecommand \Eprint[0]{\href }%
\@ifxundefined \urlstyle {%
  \providecommand \doi [1]{doi:\discretionary{}{}{}#1}%
}{%
  \providecommand \doi [0]{doi:\discretionary{}{}{}\begingroup
  \urlstyle{rm}\Url }%
}%
\providecommand \doibase [0]{http://dx.doi.org/}%
\providecommand \Doi[1]{\href{\doibase#1}}%
\providecommand \bibAnnote [3]{%
  \BibitemShut{#1}%
  \begin{quotation}\noindent
    \textsc{Key:}\ #2\\\textsc{Annotation:}\ #3%
  \end{quotation}%
}%
\providecommand \bibAnnoteFile [2]{%
  \IfFileExists{#2}{\bibAnnote {#1} {#2} {\input{#2}}}{}%
}%
\providecommand \typeout [0]{\immediate \write \m@ne }%
\providecommand \selectlanguage [0]{\@gobble}%
\providecommand \bibinfo [0]{\@secondoftwo}%
\providecommand \bibfield [0]{\@secondoftwo}%
\providecommand \translation [1]{[#1]}%
\providecommand \BibitemOpen[0]{}%
\providecommand \bibitemStop [0]{}%
\providecommand \bibitemNoStop [0]{.\EOS\space}%
\providecommand \EOS [0]{\spacefactor3000\relax}%
\providecommand \BibitemShut [1]{\csname bibitem#1\endcsname}%
%</preamble>
\bibitem{wong2004anomalous}%
  \BibitemOpen
  \bibfield{author}{%
  \bibinfo {author} {\bibfnamefont{I.}~\bibnamefont{Wong}}, \bibinfo {author}
  {\bibfnamefont{M.}~\bibnamefont{Gardel}}, \bibinfo {author}
  {\bibfnamefont{D.}~\bibnamefont{Reichman}}, \bibinfo {author}
  {\bibfnamefont{E.~R.}\ \bibnamefont{Weeks}}, \bibinfo {author}
  {\bibfnamefont{M.}~\bibnamefont{Valentine}}, \bibinfo {author}
  {\bibfnamefont{A.}~\bibnamefont{Bausch}},\ and\ \bibinfo {author}
  {\bibfnamefont{D.}~\bibnamefont{Weitz}},\ }%
  \bibfield{journal}{%
  \bibinfo {journal} {Phys. Rev. Lett.}\ }%
  \textbf{\bibinfo {volume} {92}},\ \bibinfo {pages} {178101} (\bibinfo {year}
  {2004})%
  \bibAnnoteFile{NoStop}{wong2004anomalous}%
\bibitem{wang2009anomalous}%
  \BibitemOpen
  \bibfield{author}{%
  \bibinfo {author} {\bibfnamefont{B.}~\bibnamefont{Wang}}, \bibinfo {author}
  {\bibfnamefont{S.}~\bibnamefont{Anthony}}, \bibinfo {author}
  {\bibfnamefont{S.}~\bibnamefont{Bae}},\ and\ \bibinfo {author}
  {\bibfnamefont{S.}~\bibnamefont{Granick}},\ }%
  \bibfield{journal}{%
  \bibinfo {journal} {Proc. Natl. Acad. Sci. USA}\ }%
  \textbf{\bibinfo {volume} {106}},\ \bibinfo {pages} {15160} (\bibinfo {year}
  {2009})%
  \bibAnnoteFile{NoStop}{wang2009anomalous}%
\bibitem{parry2014bacterial}%
  \BibitemOpen
  \bibfield{author}{%
  \bibinfo {author} {\bibfnamefont{B.~R.}\ \bibnamefont{Parry}}, \bibinfo
  {author} {\bibfnamefont{I.~V.}\ \bibnamefont{Surovtsev}}, \bibinfo {author}
  {\bibfnamefont{M.~T.}\ \bibnamefont{Cabeen}}, \bibinfo {author}
  {\bibfnamefont{C.~S.}\ \bibnamefont{O{'}Hern}}, \bibinfo {author}
  {\bibfnamefont{E.~R.}\ \bibnamefont{Dufresne}},\ and\ \bibinfo {author}
  {\bibfnamefont{C.}~\bibnamefont{Jacobs-Wagner}},\ }%
  \bibfield{journal}{%
  \bibinfo {journal} {Cell}\ }%
  \textbf{\bibinfo {volume} {156}},\ \bibinfo {pages} {183} (\bibinfo {year}
  {2014})%
  \bibAnnoteFile{NoStop}{parry2014bacterial}%
\bibitem{sako2000single}%
  \BibitemOpen
  \bibfield{author}{%
  \bibinfo {author} {\bibfnamefont{Y.}~\bibnamefont{Sako}}, \bibinfo {author}
  {\bibfnamefont{S.}~\bibnamefont{Minoghchi}},\ and\ \bibinfo {author}
  {\bibfnamefont{T.}~\bibnamefont{Yanagida}},\ }%
  \bibfield{journal}{%
  \bibinfo {journal} {Nat. Cell Biol.}\ }%
  \textbf{\bibinfo {volume} {2}},\ \bibinfo {pages} {168} (\bibinfo {year}
  {2000})%
  \bibAnnoteFile{NoStop}{sako2000single}%
\bibitem{cai2009single}%
  \BibitemOpen
  \bibfield{author}{%
  \bibinfo {author} {\bibfnamefont{D.}~\bibnamefont{Cai}}, \bibinfo {author}
  {\bibfnamefont{D.~P.}\ \bibnamefont{McEwen}}, \bibinfo {author}
  {\bibfnamefont{J.~R.}\ \bibnamefont{Martens}}, \bibinfo {author}
  {\bibfnamefont{E.}~\bibnamefont{Meyhofer}},\ and\ \bibinfo {author}
  {\bibfnamefont{K.~J.}\ \bibnamefont{Verhey}},\ }%
  \bibfield{journal}{%
  \bibinfo {journal} {PLoS Biol.}\ }%
  \textbf{\bibinfo {volume} {7}},\ \bibinfo {pages} {e1000216} (\bibinfo {year}
  {2009})%
  \bibAnnoteFile{NoStop}{cai2009single}%
\bibitem{zimmerman1991estimation}%
  \BibitemOpen
  \bibfield{author}{%
  \bibinfo {author} {\bibfnamefont{S.~B.}\ \bibnamefont{Zimmerman}}\ and\
  \bibinfo {author} {\bibfnamefont{S.~O.}\ \bibnamefont{Trach}},\ }%
  \bibfield{journal}{%
  \bibinfo {journal} {J. Mol. Biol.}\ }%
  \textbf{\bibinfo {volume} {222}},\ \bibinfo {pages} {599} (\bibinfo {year}
  {1991})%
  \bibAnnoteFile{NoStop}{zimmerman1991estimation}%
\bibitem{konopka2006crowding}%
  \BibitemOpen
  \bibfield{author}{%
  \bibinfo {author} {\bibfnamefont{M.~C.}\ \bibnamefont{Konopka}}, \bibinfo
  {author} {\bibfnamefont{I.~A.}\ \bibnamefont{Shkel}}, \bibinfo {author}
  {\bibfnamefont{S.}~\bibnamefont{Cayley}}, \bibinfo {author}
  {\bibfnamefont{M.~T.}\ \bibnamefont{Record}},\ and\ \bibinfo {author}
  {\bibfnamefont{J.~C.}\ \bibnamefont{Weisshaar}},\ }%
  \bibfield{journal}{%
  \bibinfo {journal} {J. Bacteriol.}\ }%
  \textbf{\bibinfo {volume} {188}},\ \bibinfo {pages} {6115} (\bibinfo {year}
  {2006})%
  \bibAnnoteFile{NoStop}{konopka2006crowding}%
\bibitem{hofling2013anomalous}%
  \BibitemOpen
  \bibfield{author}{%
  \bibinfo {author} {\bibfnamefont{F.}~\bibnamefont{H{\"o}fling}}\ and\
  \bibinfo {author} {\bibfnamefont{T.}~\bibnamefont{Franosch}},\ }%
  \bibfield{journal}{%
  \bibinfo {journal} {Rep. Prog. Phys.}\ }%
  \textbf{\bibinfo {volume} {76}},\ \bibinfo {pages} {046602} (\bibinfo {year}
  {2013})%
  \bibAnnoteFile{NoStop}{hofling2013anomalous}%
\bibitem{tabei2013intracellular}%
  \BibitemOpen
  \bibfield{author}{%
  \bibinfo {author} {\bibfnamefont{S.~A.}\ \bibnamefont{Tabei}}, \bibinfo
  {author} {\bibfnamefont{S.}~\bibnamefont{Burov}}, \bibinfo {author}
  {\bibfnamefont{H.~Y.}\ \bibnamefont{Kim}}, \bibinfo {author}
  {\bibfnamefont{A.}~\bibnamefont{Kuznetsov}}, \bibinfo {author}
  {\bibfnamefont{T.}~\bibnamefont{Huynh}}, \bibinfo {author}
  {\bibfnamefont{J.}~\bibnamefont{Jureller}}, \bibinfo {author}
  {\bibfnamefont{L.~H.}\ \bibnamefont{Philipson}}, \bibinfo {author}
  {\bibfnamefont{A.~R.}\ \bibnamefont{Dinner}},\ and\ \bibinfo {author}
  {\bibfnamefont{N.~F.}\ \bibnamefont{Scherer}},\ }%
  \bibfield{journal}{%
  \bibinfo {journal} {Proc. Natl. Acad. Sci. USA}\ }%
  \textbf{\bibinfo {volume} {110}},\ \bibinfo {pages} {4911} (\bibinfo {year}
  {2013})%
  \bibAnnoteFile{NoStop}{tabei2013intracellular}%
\bibitem{burov2013distribution}%
  \BibitemOpen
  \bibfield{author}{%
  \bibinfo {author} {\bibfnamefont{S.}~\bibnamefont{Burov}}, \bibinfo {author}
  {\bibfnamefont{S.~A.}\ \bibnamefont{Tabei}}, \bibinfo {author}
  {\bibfnamefont{T.}~\bibnamefont{Huynh}}, \bibinfo {author}
  {\bibfnamefont{M.~P.}\ \bibnamefont{Murrell}}, \bibinfo {author}
  {\bibfnamefont{L.~H.}\ \bibnamefont{Philipson}}, \bibinfo {author}
  {\bibfnamefont{S.~A.}\ \bibnamefont{Rice}}, \bibinfo {author}
  {\bibfnamefont{M.~L.}\ \bibnamefont{Gardel}}, \bibinfo {author}
  {\bibfnamefont{N.~F.}\ \bibnamefont{Scherer}},\ and\ \bibinfo {author}
  {\bibfnamefont{A.~R.}\ \bibnamefont{Dinner}},\ }%
  \bibfield{journal}{%
  \bibinfo {journal} {Proc. Natl. Acad. Sci. USA}\ }%
  \textbf{\bibinfo {volume} {110}},\ \bibinfo {pages} {19689} (\bibinfo {year}
  {2013})%
  \bibAnnoteFile{NoStop}{burov2013distribution}%
\bibitem{CoxPRL}%
  \BibitemOpen
  \bibfield{author}{%
  \bibinfo {author} {\bibfnamefont{I.}~\bibnamefont{Golding}}\ and\ \bibinfo
  {author} {\bibfnamefont{E.~C.}\ \bibnamefont{Cox}},\ }%
  \bibfield{journal}{%
  \bibinfo {journal} {Phys. Rev. Lett.}\ }%
  \textbf{\bibinfo {volume} {96}},\ \bibinfo {pages} {098102} (\bibinfo {year}
  {2006})%
  \bibAnnoteFile{NoStop}{CoxPRL}%
\bibitem{tolic2004anomalous}%
  \BibitemOpen
  \bibfield{author}{%
  \bibinfo {author} {\bibnamefont{{Toli{{\'c}}-Norrelykke, Iva Marija and
  Munteanu, Emilia-Laura and Thon, Genevieve and Oddershede, Lene and
  Berg-Sorensen, Kirstine}}},\ }%
  \bibfield{journal}{%
  \bibinfo {journal} {Phys. Rev. Lett.}\ }%
  \textbf{\bibinfo {volume} {93}},\ \bibinfo {pages} {078102} (\bibinfo {year}
  {2004})%
  \bibAnnoteFile{NoStop}{tolic2004anomalous}%
\bibitem{weber2010bacterial}%
  \BibitemOpen
  \bibfield{author}{%
  \bibinfo {author} {\bibfnamefont{S.}~\bibnamefont{Weber}}, \bibinfo {author}
  {\bibfnamefont{A.}~\bibnamefont{Spakowitz}},\ and\ \bibinfo {author}
  {\bibfnamefont{J.}~\bibnamefont{Theriot}},\ }%
  \bibfield{journal}{%
  \bibinfo {journal} {Phys. Rev. Lett.}\ }%
  \textbf{\bibinfo {volume} {104}},\ \bibinfo {pages} {238102} (\bibinfo {year}
  {2010})%
  \bibAnnoteFile{NoStop}{weber2010bacterial}%
\bibitem{weigel2011ergodic}%
  \BibitemOpen
  \bibfield{author}{%
  \bibinfo {author} {\bibfnamefont{A.}~\bibnamefont{Weigel}}, \bibinfo {author}
  {\bibfnamefont{B.}~\bibnamefont{Simon}}, \bibinfo {author}
  {\bibfnamefont{M.}~\bibnamefont{Tamkun}},\ and\ \bibinfo {author}
  {\bibfnamefont{D.}~\bibnamefont{Krapf}},\ }%
  \bibfield{journal}{%
  \bibinfo {journal} {Proc. Natl. Acad. Sci. USA}\ }%
  \textbf{\bibinfo {volume} {108}},\ \bibinfo {pages} {6438} (\bibinfo {year}
  {2011})%
  \bibAnnoteFile{NoStop}{weigel2011ergodic}%
\bibitem{bressloff2013stochastic}%
  \BibitemOpen
  \bibfield{author}{%
  \bibinfo {author} {\bibfnamefont{P.~C.}\ \bibnamefont{Bressloff}}\ and\
  \bibinfo {author} {\bibfnamefont{J.~M.}\ \bibnamefont{Newby}},\ }%
  \bibfield{journal}{%
  \bibinfo {journal} {Rev. Mod. Phys.}\ }%
  \textbf{\bibinfo {volume} {85}},\ \bibinfo {pages} {135} (\bibinfo {year}
  {2013})%
  \bibAnnoteFile{NoStop}{bressloff2013stochastic}%
\bibitem{cugliandolo1994evidence}%
  \BibitemOpen
  \bibfield{author}{%
  \bibinfo {author} {\bibfnamefont{L.}~\bibnamefont{Cugliandolo}}, \bibinfo
  {author} {\bibfnamefont{J.}~\bibnamefont{Kurchan}},\ and\ \bibinfo {author}
  {\bibfnamefont{F.}~\bibnamefont{Ritort}},\ }%
  \bibfield{journal}{%
  \bibinfo {journal} {Phys. Rev. B}\ }%
  \textbf{\bibinfo {volume} {49}},\ \bibinfo {pages} {6331} (\bibinfo {year}
  {1994})%
  \bibAnnoteFile{NoStop}{cugliandolo1994evidence}%
\bibitem{manzo2015weak}%
  \BibitemOpen
  \bibfield{author}{%
  \bibinfo {author} {\bibfnamefont{C.}~\bibnamefont{Manzo}}, \bibinfo {author}
  {\bibfnamefont{J.~A.}\ \bibnamefont{Torreno-Pina}}, \bibinfo {author}
  {\bibfnamefont{P.}~\bibnamefont{Massignan}}, \bibinfo {author}
  {\bibfnamefont{G.~J.}\ \bibnamefont{Lapeyre~Jr}}, \bibinfo {author}
  {\bibfnamefont{M.}~\bibnamefont{Lewenstein}},\ and\ \bibinfo {author}
  {\bibfnamefont{M.~F.~G.}\ \bibnamefont{Parajo}},\ }%
  \bibfield{journal}{%
  \bibinfo {journal} {PRX}\ }%
  \textbf{\bibinfo {volume} {5}},\ \bibinfo {pages} {011021} (\bibinfo {year}
  {2015})%
  \bibAnnoteFile{NoStop}{manzo2015weak}%
\bibitem{Seino2011}%
  \BibitemOpen
  \bibfield{author}{%
  \bibinfo {author} {\bibfnamefont{S.}~\bibnamefont{Seino}}, \bibinfo {author}
  {\bibfnamefont{T.}~\bibnamefont{Shibasaki}},\ and\ \bibinfo {author}
  {\bibfnamefont{K.}~\bibnamefont{Minami}},\ }%
  \bibfield{journal}{%
  \bibinfo {journal} {J. Clin. Invest.}\ }%
  \textbf{\bibinfo {volume} {121}},\ \bibinfo {pages} {2118} (\bibinfo {year}
  {2011})%
  \bibAnnoteFile{NoStop}{Seino2011}%
\bibitem{bouchaud1995aging}%
  \BibitemOpen
  \bibfield{author}{%
  \bibinfo {author} {\bibfnamefont{J.-P.}\ \bibnamefont{Bouchaud}}\ and\
  \bibinfo {author} {\bibfnamefont{D.~S.}\ \bibnamefont{Dean}},\ }%
  \bibfield{journal}{%
  \bibinfo {journal} {J. Phys. I}\ }%
  \textbf{\bibinfo {volume} {5}},\ \bibinfo {pages} {265} (\bibinfo {year}
  {1995})%
  \bibAnnoteFile{NoStop}{bouchaud1995aging}%
\bibitem{feigel1988stochastic}%
  \BibitemOpen
  \bibfield{author}{%
  \bibinfo {author} {\bibfnamefont{M.}~\bibnamefont{Feigel'Man}}\ and\ \bibinfo
  {author} {\bibfnamefont{V.}~\bibnamefont{Vinokur}},\ }%
  \bibfield{journal}{%
  \bibinfo {journal} {J. Phys. Paris}\ }%
  \textbf{\bibinfo {volume} {49}},\ \bibinfo {pages} {1731} (\bibinfo {year}
  {1988})%
  \bibAnnoteFile{NoStop}{feigel1988stochastic}%
\bibitem{Dix2008}%
  \BibitemOpen
  \bibfield{author}{%
  \bibinfo {author} {\bibfnamefont{J.~A.}\ \bibnamefont{Dix}}\ and\ \bibinfo
  {author} {\bibfnamefont{A.~S.}\ \bibnamefont{Verkman}},\ }%
  \bibfield{journal}{%
  \bibinfo {journal} {Annu. Rev. Biophys.}\ }%
  \textbf{\bibinfo {volume} {37}},\ \bibinfo {pages} {247} (\bibinfo {year}
  {2008})%
  \bibAnnoteFile{NoStop}{Dix2008}%
\bibitem{varadi2003kinesin}%
  \BibitemOpen
  \bibfield{author}{%
  \bibinfo {author} {\bibfnamefont{A.}~\bibnamefont{Varadi}}, \bibinfo {author}
  {\bibfnamefont{T.}~\bibnamefont{Tsuboi}}, \bibinfo {author}
  {\bibfnamefont{L.~I.}\ \bibnamefont{Johnson-Cadwell}}, \bibinfo {author}
  {\bibfnamefont{V.~J.}\ \bibnamefont{Allan}},\ and\ \bibinfo {author}
  {\bibfnamefont{G.~A.}\ \bibnamefont{Rutter}},\ }%
  \bibfield{journal}{%
  \bibinfo {journal} {Biochem. Biophys. Res. Commun.}\ }%
  \textbf{\bibinfo {volume} {311}},\ \bibinfo {pages} {272} (\bibinfo {year}
  {2003})%
  \bibAnnoteFile{NoStop}{varadi2003kinesin}%
\bibitem{vale1992directional}%
  \BibitemOpen
  \bibfield{author}{%
  \bibinfo {author} {\bibfnamefont{R.~D.}\ \bibnamefont{Vale}}, \bibinfo
  {author} {\bibfnamefont{F.}~\bibnamefont{Malik}},\ and\ \bibinfo {author}
  {\bibfnamefont{D.}~\bibnamefont{Brown}},\ }%
  \bibfield{journal}{%
  \bibinfo {journal} {J. Cell Biol.}\ }%
  \textbf{\bibinfo {volume} {119}},\ \bibinfo {pages} {1589} (\bibinfo {year}
  {1992})%
  \bibAnnoteFile{NoStop}{vale1992directional}%
\bibitem{muller2008tug}%
  \BibitemOpen
  \bibfield{author}{%
  \bibinfo {author} {\bibfnamefont{M.~J.}\ \bibnamefont{M{\"u}ller}}, \bibinfo
  {author} {\bibfnamefont{S.}~\bibnamefont{Klumpp}},\ and\ \bibinfo {author}
  {\bibfnamefont{R.}~\bibnamefont{Lipowsky}},\ }%
  \bibfield{journal}{%
  \bibinfo {journal} {Proc. Natl. Acad. Sci. USA}\ }%
  \textbf{\bibinfo {volume} {105}},\ \bibinfo {pages} {4609} (\bibinfo {year}
  {2008})%
  \bibAnnoteFile{NoStop}{muller2008tug}%
\bibitem{newby2010local}%
  \BibitemOpen
  \bibfield{author}{%
  \bibinfo {author} {\bibfnamefont{J.}~\bibnamefont{Newby}}\ and\ \bibinfo
  {author} {\bibfnamefont{P.~C.}\ \bibnamefont{Bressloff}},\ }%
  \bibfield{journal}{%
  \bibinfo {journal} {Phys. Biol.}\ }%
  \textbf{\bibinfo {volume} {7}},\ \bibinfo {pages} {036004} (\bibinfo {year}
  {2010})%
  \bibAnnoteFile{NoStop}{newby2010local}%
\bibitem{hendricks2010motor}%
  \BibitemOpen
  \bibfield{author}{%
  \bibinfo {author} {\bibfnamefont{A.~G.}\ \bibnamefont{Hendricks}}, \bibinfo
  {author} {\bibfnamefont{E.}~\bibnamefont{Perlson}}, \bibinfo {author}
  {\bibfnamefont{J.~L.}\ \bibnamefont{Ross}}, \bibinfo {author}
  {\bibfnamefont{H.~W.}\ \bibnamefont{Schroeder~III}}, \bibinfo {author}
  {\bibfnamefont{M.}~\bibnamefont{Tokito}},\ and\ \bibinfo {author}
  {\bibfnamefont{E.~L.}\ \bibnamefont{Holzbaur}},\ }%
  \bibfield{journal}{%
  \bibinfo {journal} {Curr. Biol.}\ }%
  \textbf{\bibinfo {volume} {20}},\ \bibinfo {pages} {697} (\bibinfo {year}
  {2010})%
  \bibAnnoteFile{NoStop}{hendricks2010motor}%
\bibitem{soppina2009tug}%
  \BibitemOpen
  \bibfield{author}{%
  \bibinfo {author} {\bibfnamefont{V.}~\bibnamefont{Soppina}}, \bibinfo
  {author} {\bibfnamefont{A.~K.}\ \bibnamefont{Rai}}, \bibinfo {author}
  {\bibfnamefont{A.~J.}\ \bibnamefont{Ramaiya}}, \bibinfo {author}
  {\bibfnamefont{P.}~\bibnamefont{Barak}},\ and\ \bibinfo {author}
  {\bibfnamefont{R.}~\bibnamefont{Mallik}},\ }%
  \bibfield{journal}{%
  \bibinfo {journal} {Proc. Natl. Acad. Sci. USA}\ }%
  \textbf{\bibinfo {volume} {106}},\ \bibinfo {pages} {19381} (\bibinfo {year}
  {2009})%
  \bibAnnoteFile{NoStop}{soppina2009tug}%
\bibitem{ali2011myosin}%
  \BibitemOpen
  \bibfield{author}{%
  \bibinfo {author} {\bibfnamefont{M.~Y.}\ \bibnamefont{Ali}}, \bibinfo
  {author} {\bibfnamefont{G.~G.}\ \bibnamefont{Kennedy}}, \bibinfo {author}
  {\bibfnamefont{D.}~\bibnamefont{Safer}}, \bibinfo {author}
  {\bibfnamefont{K.~M.}\ \bibnamefont{Trybus}}, \bibinfo {author}
  {\bibfnamefont{H.~L.}\ \bibnamefont{Sweeney}},\ and\ \bibinfo {author}
  {\bibfnamefont{D.~M.}\ \bibnamefont{Warshaw}},\ }%
  \bibfield{journal}{%
  \bibinfo {journal} {Proc. Natl. Acad. Sci. USA}\ }%
  \textbf{\bibinfo {volume} {108}},\ \bibinfo {pages} {E535} (\bibinfo {year}
  {2011})%
  \bibAnnoteFile{NoStop}{ali2011myosin}%
\bibitem{hancock2014bidirectional}%
  \BibitemOpen
  \bibfield{author}{%
  \bibinfo {author} {\bibfnamefont{W.~O.}\ \bibnamefont{Hancock}},\ }%
  \bibfield{journal}{%
  \bibinfo {journal} {Nat. Rev. Mol. Cell Biol.}\ }%
  \textbf{\bibinfo {volume} {15}},\ \bibinfo {pages} {615} (\bibinfo {year}
  {2014})%
  \bibAnnoteFile{NoStop}{hancock2014bidirectional}%
\bibitem{lee2015axonal}%
  \BibitemOpen
  \bibfield{author}{%
  \bibinfo {author} {\bibfnamefont{R.~H.}\ \bibnamefont{Lee}}\ and\ \bibinfo
  {author} {\bibfnamefont{C.~S.}\ \bibnamefont{Mitchell}},\ }%
  \bibfield{journal}{%
  \bibinfo {journal} {J. Theor. Biol.}}%
   (\bibinfo {year} {2015})%
  \bibAnnoteFile{NoStop}{lee2015axonal}%
\bibitem{finer1994single}%
  \BibitemOpen
  \bibfield{author}{%
  \bibinfo {author} {\bibfnamefont{J.~T.}\ \bibnamefont{Finer}}, \bibinfo
  {author} {\bibfnamefont{R.~M.}\ \bibnamefont{Simmons}}, \bibinfo {author}
  {\bibfnamefont{J.~A.}\ \bibnamefont{Spudich}}, \emph{et~al.},\ }%
  \bibfield{journal}{%
  \bibinfo {journal} {Nature}\ }%
  \textbf{\bibinfo {volume} {368}},\ \bibinfo {pages} {113} (\bibinfo {year}
  {1994})%
  \bibAnnoteFile{NoStop}{finer1994single}%
\bibitem{howard1997molecular}%
  \BibitemOpen
  \bibfield{author}{%
  \bibinfo {author} {\bibfnamefont{J.}~\bibnamefont{Howard}},\ }%
  \bibfield{journal}{%
  \bibinfo {journal} {Nature}\ }%
  \textbf{\bibinfo {volume} {389}},\ \bibinfo {pages} {561} (\bibinfo {year}
  {1997})%
  \bibAnnoteFile{NoStop}{howard1997molecular}%
\bibitem{Pierobon20094268}%
  \BibitemOpen
  \bibfield{author}{%
  \bibinfo {author} {\bibfnamefont{P.}~\bibnamefont{Pierobon}}, \bibinfo
  {author} {\bibfnamefont{S.}~\bibnamefont{Achouri}}, \bibinfo {author}
  {\bibfnamefont{S.}~\bibnamefont{Courty}}, \bibinfo {author}
  {\bibfnamefont{A.~R.}\ \bibnamefont{Dunn}}, \bibinfo {author}
  {\bibfnamefont{J.~A.}\ \bibnamefont{Spudich}}, \bibinfo {author}
  {\bibfnamefont{M.}~\bibnamefont{Dahan}},\ and\ \bibinfo {author}
  {\bibfnamefont{G.}~\bibnamefont{Cappello}},\ }%
  \bibfield{journal}{%
  \bibinfo {journal} {Biophys. J.}\ }%
  \textbf{\bibinfo {volume} {96}},\ \bibinfo {pages} {4268 } (\bibinfo {year}
  {2009})%
  \bibAnnoteFile{NoStop}{Pierobon20094268}%
\bibitem{saxton1997single}%
  \BibitemOpen
  \bibfield{author}{%
  \bibinfo {author} {\bibfnamefont{M.~J.}\ \bibnamefont{Saxton}}\ and\ \bibinfo
  {author} {\bibfnamefont{K.}~\bibnamefont{Jacobson}},\ }%
  \bibfield{journal}{%
  \bibinfo {journal} {Annu. Rev. Biophys. Biomol. Struct.}\ }%
  \textbf{\bibinfo {volume} {26}},\ \bibinfo {pages} {373} (\bibinfo {year}
  {1997})%
  \bibAnnoteFile{NoStop}{saxton1997single}%
\bibitem{bouchaud1992weak}%
  \BibitemOpen
  \bibfield{author}{%
  \bibinfo {author} {\bibfnamefont{J.-P.}\ \bibnamefont{Bouchaud}},\ }%
  \bibfield{journal}{%
  \bibinfo {journal} {J. Phys. I}\ }%
  \textbf{\bibinfo {volume} {2}},\ \bibinfo {pages} {1705} (\bibinfo {year}
  {1992})%
  \bibAnnoteFile{NoStop}{bouchaud1992weak}%
\bibitem{lubelski2008nonergodicity}%
  \BibitemOpen
  \bibfield{author}{%
  \bibinfo {author} {\bibfnamefont{A.}~\bibnamefont{Lubelski}}, \bibinfo
  {author} {\bibfnamefont{I.~M.}\ \bibnamefont{Sokolov}},\ and\ \bibinfo
  {author} {\bibfnamefont{J.}~\bibnamefont{Klafter}},\ }%
  \bibfield{journal}{%
  \bibinfo {journal} {Phys. Rev. Lett.}\ }%
  \textbf{\bibinfo {volume} {100}},\ \bibinfo {pages} {250602} (\bibinfo {year}
  {2008})%
  \bibAnnoteFile{NoStop}{lubelski2008nonergodicity}%
\bibitem{he2008random}%
  \BibitemOpen
  \bibfield{author}{%
  \bibinfo {author} {\bibfnamefont{Y.}~\bibnamefont{He}}, \bibinfo {author}
  {\bibfnamefont{S.}~\bibnamefont{Burov}}, \bibinfo {author}
  {\bibfnamefont{R.}~\bibnamefont{Metzler}},\ and\ \bibinfo {author}
  {\bibfnamefont{E.}~\bibnamefont{Barkai}},\ }%
  \bibfield{journal}{%
  \bibinfo {journal} {Phys. Rev. Lett.}\ }%
  \textbf{\bibinfo {volume} {101}},\ \bibinfo {pages} {058101} (\bibinfo {year}
  {2008})%
  \bibAnnoteFile{NoStop}{he2008random}%
\bibitem{sokolov2008viewpoint}%
  \BibitemOpen
  \bibfield{author}{%
  \bibinfo {author} {\bibfnamefont{I.~M.}\ \bibnamefont{Sokolov}},\ }%
  \bibfield{journal}{%
  \bibinfo {journal} {Physics}\ }%
  \textbf{\bibinfo {volume} {1}},\ \bibinfo {pages} {8} (\bibinfo {year}
  {2008})%
  \bibAnnoteFile{NoStop}{sokolov2008viewpoint}%
\bibitem{schulz2013aging}%
  \BibitemOpen
  \bibfield{author}{%
  \bibinfo {author} {\bibfnamefont{J.~H.}\ \bibnamefont{Schulz}}, \bibinfo
  {author} {\bibfnamefont{E.}~\bibnamefont{Barkai}},\ and\ \bibinfo {author}
  {\bibfnamefont{R.}~\bibnamefont{Metzler}},\ }%
  \bibfield{journal}{%
  \bibinfo {journal} {Phys. Rev. Lett.}\ }%
  \textbf{\bibinfo {volume} {110}},\ \bibinfo {pages} {020602} (\bibinfo {year}
  {2013})%
  \bibAnnoteFile{NoStop}{schulz2013aging}%
\bibitem{solomon1993observation}%
  \BibitemOpen
  \bibfield{author}{%
  \bibinfo {author} {\bibfnamefont{T.}~\bibnamefont{Solomon}}, \bibinfo
  {author} {\bibfnamefont{E.~R.}\ \bibnamefont{Weeks}},\ and\ \bibinfo {author}
  {\bibfnamefont{H.~L.}\ \bibnamefont{Swinney}},\ }%
  \bibfield{journal}{%
  \bibinfo {journal} {Phys. Rev. Lett.}\ }%
  \textbf{\bibinfo {volume} {71}},\ \bibinfo {pages} {3975} (\bibinfo {year}
  {1993})%
  \bibAnnoteFile{NoStop}{solomon1993observation}%
\bibitem{devoe2009power}%
  \BibitemOpen
  \bibfield{author}{%
  \bibinfo {author} {\bibfnamefont{R.~G.}\ \bibnamefont{DeVoe}},\ }%
  \bibfield{journal}{%
  \bibinfo {journal} {Phys. Rev. Lett.}\ }%
  \textbf{\bibinfo {volume} {102}},\ \bibinfo {pages} {063001} (\bibinfo
  {month} {Feb}\ \bibinfo {year} {2009})%
  \bibAnnoteFile{NoStop}{devoe2009power}%
\bibitem{lundh2000hydrodynamic}%
  \BibitemOpen
  \bibfield{author}{%
  \bibinfo {author} {\bibfnamefont{E.}~\bibnamefont{Lundh}}\ and\ \bibinfo
  {author} {\bibfnamefont{P.}~\bibnamefont{Ao}},\ }%
  \bibfield{journal}{%
  \bibinfo {journal} {Phys. Rev. A}\ }%
  \textbf{\bibinfo {volume} {61}},\ \bibinfo {pages} {063612} (\bibinfo {year}
  {2000})%
  \bibAnnoteFile{NoStop}{lundh2000hydrodynamic}%
\bibitem{shraiman1987diffusive}%
  \BibitemOpen
  \bibfield{author}{%
  \bibinfo {author} {\bibfnamefont{B.~I.}\ \bibnamefont{Shraiman}},\ }%
  \bibfield{journal}{%
  \bibinfo {journal} {Phys. Rev. A}\ }%
  \textbf{\bibinfo {volume} {36}},\ \bibinfo {pages} {261} (\bibinfo {year}
  {1987})%
  \bibAnnoteFile{NoStop}{shraiman1987diffusive}%
\bibitem{thoresen2013thick}%
  \BibitemOpen
  \bibfield{author}{%
  \bibinfo {author} {\bibfnamefont{T.}~\bibnamefont{Thoresen}}, \bibinfo
  {author} {\bibfnamefont{M.}~\bibnamefont{Lenz}},\ and\ \bibinfo {author}
  {\bibfnamefont{M.~L.}\ \bibnamefont{Gardel}},\ }%
  \bibfield{journal}{%
  \bibinfo {journal} {Biophysical Journal}\ }%
  \textbf{\bibinfo {volume} {104}},\ \bibinfo {pages} {655} (\bibinfo {year}
  {2013})%
  \bibAnnoteFile{NoStop}{thoresen2013thick}%
\bibitem{daniel_b_allan_2014_9971}%
  \BibitemOpen
  \bibfield{author}{%
  \bibinfo {author} {\bibfnamefont{D.~B.}\ \bibnamefont{Allan}}, \bibinfo
  {author} {\bibfnamefont{T.~A.}\ \bibnamefont{Caswell}},\ and\ \bibinfo
  {author} {\bibfnamefont{N.~C.}\ \bibnamefont{Keim}},\ }%
  \enquote{\bibinfo {title} {{Trackpy v0.2}},}\  (\bibinfo {month} {May}\
  \bibinfo {year} {2014}),\ \url{http://dx.doi.org/10.5281/zenodo.9971}%
  \bibAnnoteFile{NoStop}{daniel_b_allan_2014_9971}%
\bibitem{weiss2004anomalous}%
  \BibitemOpen
  \bibfield{author}{%
  \bibinfo {author} {\bibfnamefont{M.}~\bibnamefont{Weiss}}, \bibinfo {author}
  {\bibfnamefont{M.}~\bibnamefont{Elsner}}, \bibinfo {author}
  {\bibfnamefont{F.}~\bibnamefont{Kartberg}},\ and\ \bibinfo {author}
  {\bibfnamefont{T.}~\bibnamefont{Nilsson}},\ }%
  \bibfield{journal}{%
  \bibinfo {journal} {Biophys. J.}\ }%
  \textbf{\bibinfo {volume} {87}},\ \bibinfo {pages} {3518} (\bibinfo {year}
  {2004})%
  \bibAnnoteFile{NoStop}{weiss2004anomalous}%
\bibitem{amblard1996subdiffusion}%
  \BibitemOpen
  \bibfield{author}{%
  \bibinfo {author} {\bibfnamefont{F.}~\bibnamefont{Amblard}}, \bibinfo
  {author} {\bibfnamefont{A.~C.}\ \bibnamefont{Maggs}}, \bibinfo {author}
  {\bibfnamefont{B.}~\bibnamefont{Yurke}}, \bibinfo {author}
  {\bibfnamefont{A.~N.}\ \bibnamefont{Pargellis}},\ and\ \bibinfo {author}
  {\bibfnamefont{S.}~\bibnamefont{Leibler}},\ }%
  \bibfield{journal}{%
  \bibinfo {journal} {Phys. Rev. Lett.}\ }%
  \textbf{\bibinfo {volume} {77}},\ \bibinfo {pages} {4470} (\bibinfo {year}
  {1996})%
  \bibAnnoteFile{NoStop}{amblard1996subdiffusion}%
\bibitem{jeon2011vivo}%
  \BibitemOpen
  \bibfield{author}{%
  \bibinfo {author} {\bibfnamefont{J.-H.}\ \bibnamefont{Jeon}}, \bibinfo
  {author} {\bibfnamefont{V.}~\bibnamefont{Tejedor}}, \bibinfo {author}
  {\bibfnamefont{S.}~\bibnamefont{Burov}}, \bibinfo {author}
  {\bibfnamefont{E.}~\bibnamefont{Barkai}}, \bibinfo {author}
  {\bibfnamefont{C.}~\bibnamefont{Selhuber-Unkel}}, \bibinfo {author}
  {\bibfnamefont{K.}~\bibnamefont{Berg-S{\o}rensen}}, \bibinfo {author}
  {\bibfnamefont{L.}~\bibnamefont{Oddershede}},\ and\ \bibinfo {author}
  {\bibfnamefont{R.}~\bibnamefont{Metzler}},\ }%
  \bibfield{journal}{%
  \bibinfo {journal} {Phys. Rev. Lett.}\ }%
  \textbf{\bibinfo {volume} {106}},\ \bibinfo {pages} {048103} (\bibinfo {year}
  {2011})%
  \bibAnnoteFile{NoStop}{jeon2011vivo}%
\bibitem{reilein2005self}%
  \BibitemOpen
  \bibfield{author}{%
  \bibinfo {author} {\bibfnamefont{A.}~\bibnamefont{Reilein}}, \bibinfo
  {author} {\bibfnamefont{S.}~\bibnamefont{Yamada}},\ and\ \bibinfo {author}
  {\bibfnamefont{W.~J.}\ \bibnamefont{Nelson}},\ }%
  \bibfield{journal}{%
  \bibinfo {journal} {J. Cell Biol.}\ }%
  \textbf{\bibinfo {volume} {171}},\ \bibinfo {pages} {845} (\bibinfo {year}
  {2005})%
  \bibAnnoteFile{NoStop}{reilein2005self}%
\bibitem{ross2008kinesin}%
  \BibitemOpen
  \bibfield{author}{%
  \bibinfo {author} {\bibfnamefont{J.~L.}\ \bibnamefont{Ross}}, \bibinfo
  {author} {\bibfnamefont{H.}~\bibnamefont{Shuman}}, \bibinfo {author}
  {\bibfnamefont{E.~L.}\ \bibnamefont{Holzbaur}},\ and\ \bibinfo {author}
  {\bibfnamefont{Y.~E.}\ \bibnamefont{Goldman}},\ }%
  \bibfield{journal}{%
  \bibinfo {journal} {Biophys. J}\ }%
  \textbf{\bibinfo {volume} {94}},\ \bibinfo {pages} {3115} (\bibinfo {year}
  {2008})%
  \bibAnnoteFile{NoStop}{ross2008kinesin}%
\bibitem{osunbayo2015cargo}%
  \BibitemOpen
  \bibfield{author}{%
  \bibinfo {author} {\bibfnamefont{O.}~\bibnamefont{Osunbayo}}, \bibinfo
  {author} {\bibfnamefont{J.}~\bibnamefont{Butterfield}}, \bibinfo {author}
  {\bibfnamefont{J.}~\bibnamefont{Bergman}}, \bibinfo {author}
  {\bibfnamefont{L.}~\bibnamefont{Mershon}}, \bibinfo {author}
  {\bibfnamefont{V.}~\bibnamefont{Rodionov}},\ and\ \bibinfo {author}
  {\bibfnamefont{M.}~\bibnamefont{Vershinin}},\ }%
  \bibfield{journal}{%
  \bibinfo {journal} {Biophys. J.}\ }%
  \textbf{\bibinfo {volume} {108}},\ \bibinfo {pages} {1480} (\bibinfo {year}
  {2015})%
  \bibAnnoteFile{NoStop}{osunbayo2015cargo}%
\bibitem{elting2012future}%
  \BibitemOpen
  \bibfield{author}{%
  \bibinfo {author} {\bibfnamefont{M.~W.}\ \bibnamefont{Elting}}\ and\ \bibinfo
  {author} {\bibfnamefont{J.~A.}\ \bibnamefont{Spudich}},\ }%
  \bibfield{journal}{%
  \bibinfo {journal} {Developmental Cell}\ }%
  \textbf{\bibinfo {volume} {23}},\ \bibinfo {pages} {1084} (\bibinfo {year}
  {2012})%
  \bibAnnoteFile{NoStop}{elting2012future}%
\end{thebibliography}

\begin{thebibliography}{52}%
\makeatletter
\providecommand \@ifxundefined [1]{%
 \@ifx{#1\undefined}
}%
\providecommand \@ifnum [1]{%
 \ifnum #1\expandafter \@firstoftwo
 \else \expandafter \@secondoftwo
 \fi
}%
\providecommand \@ifx [1]{%
 \ifx #1\expandafter \@firstoftwo
 \else \expandafter \@secondoftwo
 \fi
}%
\providecommand \natexlab [1]{#1}%
\providecommand \enquote  [1]{``#1''}%
\providecommand \bibnamefont  [1]{#1}%
\providecommand \bibfnamefont [1]{#1}%
\providecommand \citenamefont [1]{#1}%
\providecommand \href@noop [0]{\@secondoftwo}%
\providecommand \href [0]{\begingroup \@sanitize@url \@href}%
\providecommand \@href[1]{\@@startlink{#1}\@@href}%
\providecommand \@@href[1]{\endgroup#1\@@endlink}%
\providecommand \@sanitize@url [0]{\catcode `\\12\catcode `\$12\catcode
  `\&12\catcode `\#12\catcode `\^12\catcode `\_12\catcode `\%12\relax}%
\providecommand \@@startlink[1]{}%
\providecommand \@@endlink[0]{}%
\providecommand \url  [0]{\begingroup\@sanitize@url \@url }%
\providecommand \@url [1]{\endgroup\@href {#1}{\urlprefix }}%
\providecommand \urlprefix  [0]{URL }%
\providecommand \Eprint [0]{\href }%
\providecommand \doibase [0]{http://dx.doi.org/}%
\providecommand \selectlanguage [0]{\@gobble}%
\providecommand \bibinfo  [0]{\@secondoftwo}%
\providecommand \bibfield  [0]{\@secondoftwo}%
\providecommand \translation [1]{[#1]}%
\providecommand \BibitemOpen [0]{}%
\providecommand \bibitemStop [0]{}%
\providecommand \bibitemNoStop [0]{.\EOS\space}%
\providecommand \EOS [0]{\spacefactor3000\relax}%
\providecommand \BibitemShut  [1]{\csname bibitem#1\endcsname}%
\let\auto@bib@innerbib\@empty
%</preamble>
\bibitem [{\citenamefont {Burov}\ \emph {et~al.}(2013)\citenamefont {Burov},
  \citenamefont {Tabei}, \citenamefont {Huynh}, \citenamefont {Murrell},
  \citenamefont {Philipson}, \citenamefont {Rice}, \citenamefont {Gardel},
  \citenamefont {Scherer},\ and\ \citenamefont
  {Dinner}}]{burov2013distribution}%
 \BibitemOpen
  \bibfield  {author} {\bibinfo {author} {\bibfnamefont {S.}\
  \bibnamefont {Burov}}, \bibinfo {author} {\bibfnamefont {S.M.A.}\
  \bibnamefont {Tabei}}, \bibinfo {author} {\bibfnamefont {T.}\ \bibnamefont
  {Huynh}}, \bibinfo {author} {\bibfnamefont {M.P.}\ \bibnamefont
  {Murrell}}, \bibinfo {author} {\bibfnamefont {L.H.}\ \bibnamefont
  {Philipson}}, \bibinfo {author} {\bibfnamefont {S.A.}\ \bibnamefont
  {Rice}}, \bibinfo {author} {\bibfnamefont {M.L.}\ \bibnamefont
  {Gardel}}, \bibinfo {author} {\bibfnamefont {N.F.}\ \bibnamefont
  {Scherer}}, \ and\ \bibinfo {author} {\bibfnamefont {A.R.}\ \bibnamefont
  {Dinner}},\ }
  {\bibfield  {journal} {\bibinfo  {journal} {Proc. Natl. Acad. Sci. USA}\
  }\textbf {\bibinfo {volume} {110}},\ \bibinfo {pages} {19689--19694}
  (\bibinfo {year} {2013})}\BibitemShut {NoStop}%
\end{thebibliography}

%Merlin.mbs v4.21 2009-07-09.
%

\clearpage

\renewcommand{\thefigure}{S\arabic{figure}}

\setcounter{figure}{0}
\begin{center}
{\bf Supplementary Information for\\A cycling state that can lead to glassy dynamics in intracellular transport}
\end{center}

\section{Alternative models}

\begin{figure}[htb]
\begin{center}
 \includegraphics[width=0.75\textwidth]{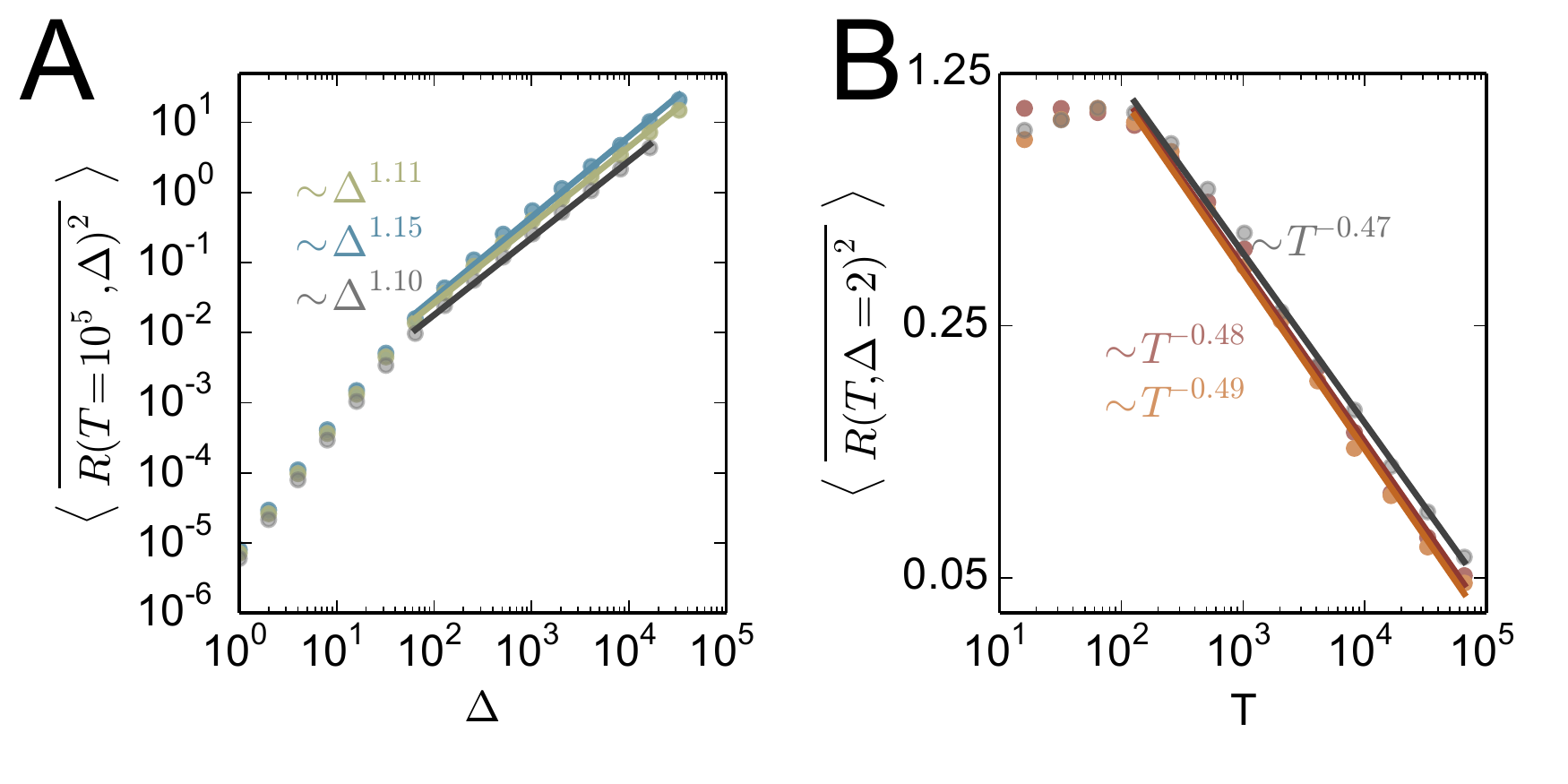}
 \end{center}
\caption{Mean-squared displacement for a Boltzmann binding scheme (colored curves), compared with results for the two-step attachment scheme in the main text (gray). (A) Time-averaged mean square displacement (MSD) for $E_0 = 0.1$ (green) and $E_0 = 0.5$ (blue).  (B) The MSD as a function of measurement time for $E_0 = 0.1$ (orange) and $E_0 = 0.5$ (red).  Other parameters are the same as those in the main text (Table 1).}
\label{fig:s1}
\end{figure}

Here we illustrate that the results are robust for alternative formulations of the model. 
First, we tested a model with a one-step binding process. In this scheme, the probability $p$ of a motor at distance $d_i$ to be attached and active is 
\begin{displaymath}
p(d_i) = \frac{\exp[-(3d_i/2s)^2]}{\exp[-E_0]+ \sum_i \exp[-(3d_i/2s)^2]}.
\end{displaymath}
Here, $E_0$ is the maximum attractive energy that can be achieved by binding, relative to the unbound state.
Fig. \ref{fig:s1} compares the MSD results for this scheme and that in the main text. 

\begin{figure}[htb]
\begin{center}
 \includegraphics[width=0.75\textwidth]{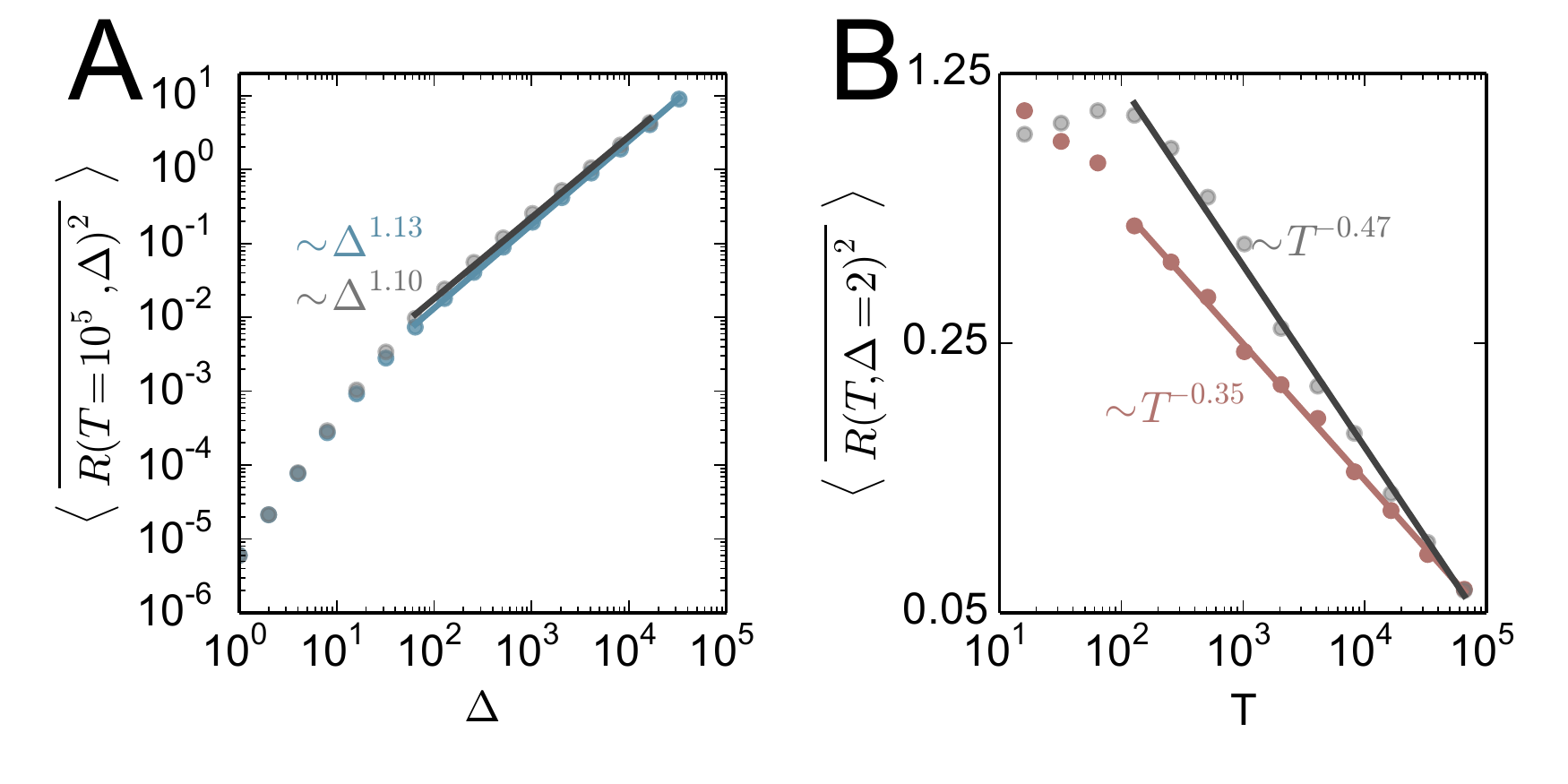}
 \end{center}
\caption{Mean-squared displacement for the model with a net velocity (red and blue).  Results for the two-step attachment scheme in the main text are shown in gray. The parameters used are the same as those in the main text (Table 1).}
\label{fig:s2}
\end{figure}

Second, we explored alternative ways of moving the motors in response to the forces.  We proposed the projection rule in the main text to ensure that motors walk parallel to tracks, as observed in many experiments. When this requirement is relaxed, the velocity at each time step can be written as $\vec{v} = \sum_i k_i  \vec{e_i}$, where $k_i $ is the number of binding sites attached to filament $i$ and $\vec{e_i} $ is the unit vector in the direction of the filament. The position update then becomes $\vec{x}(t + dt ) = \vec{v}dt$. We recover aging and similar values for the MSD scaling exponent under this relaxed assumption (Fig. \ref{fig:s2}). 

\section{Effect of nonergodicity on lag-time exponents}

\begin{figure}[htb]
\begin{center}
 \includegraphics[width=0.75\textwidth]{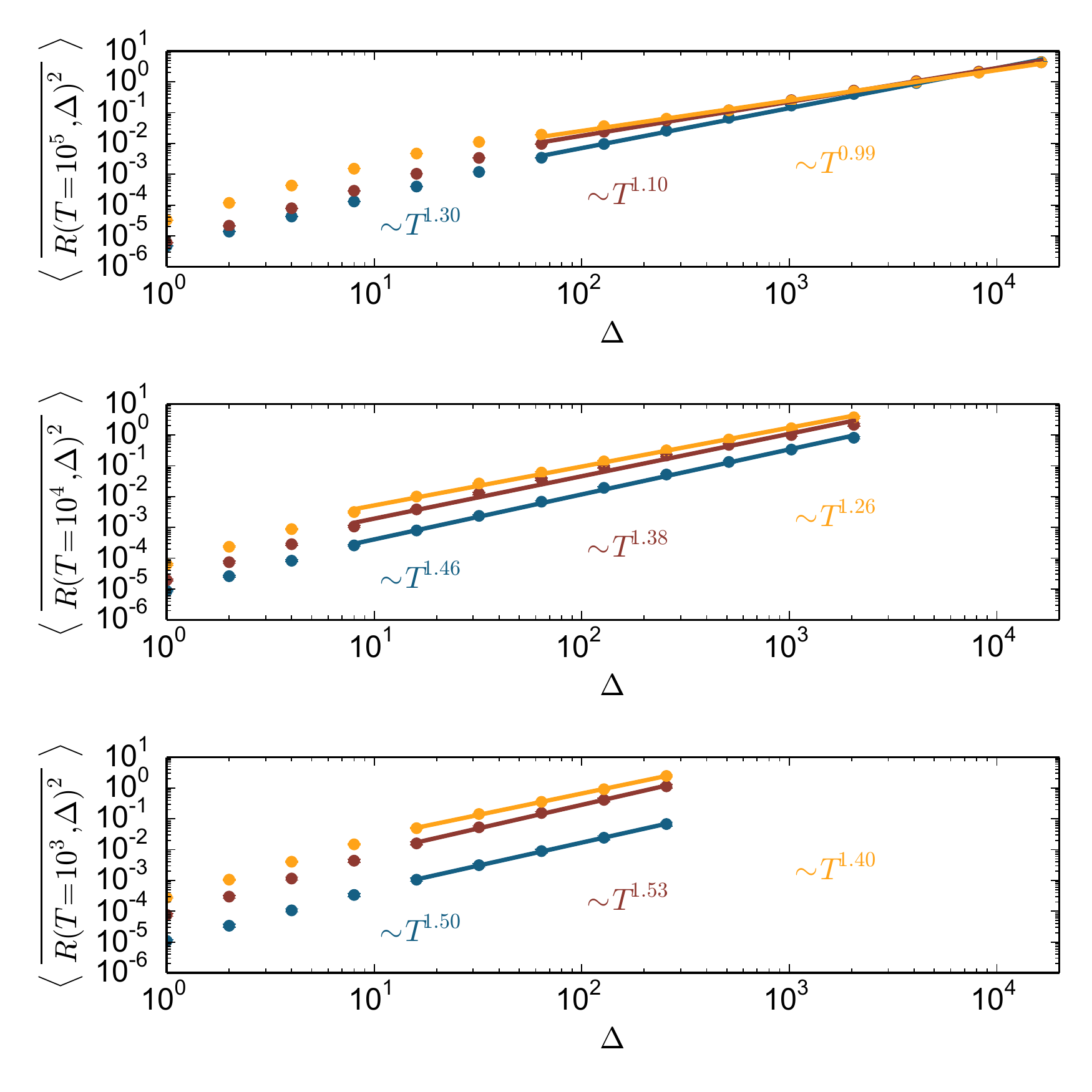}
 \end{center}
\caption{MSD as a function of lag time for binding radii $s = 0.001$ (gray), 0.01 (blue), 0.1 (orange), and 1(red) and duration of simulation $T =10^5$, $10^4$ and $10^3$ (from top to bottom).
The parameters used are the same as those in the main text (Table 1).}
\label{fig:s3}
\end{figure}

The scaling of the MSD with lag time is dependent on the duration of the recording (measurement time) $T$. Since experimental data are often limited to short trajectories, we show the resulting time-averaged MSD exponents for shorter simulation times in Fig. \ref{fig:s3}. 

\begin{figure}[htb]
\begin{center}
 \includegraphics[width=0.75\textwidth]{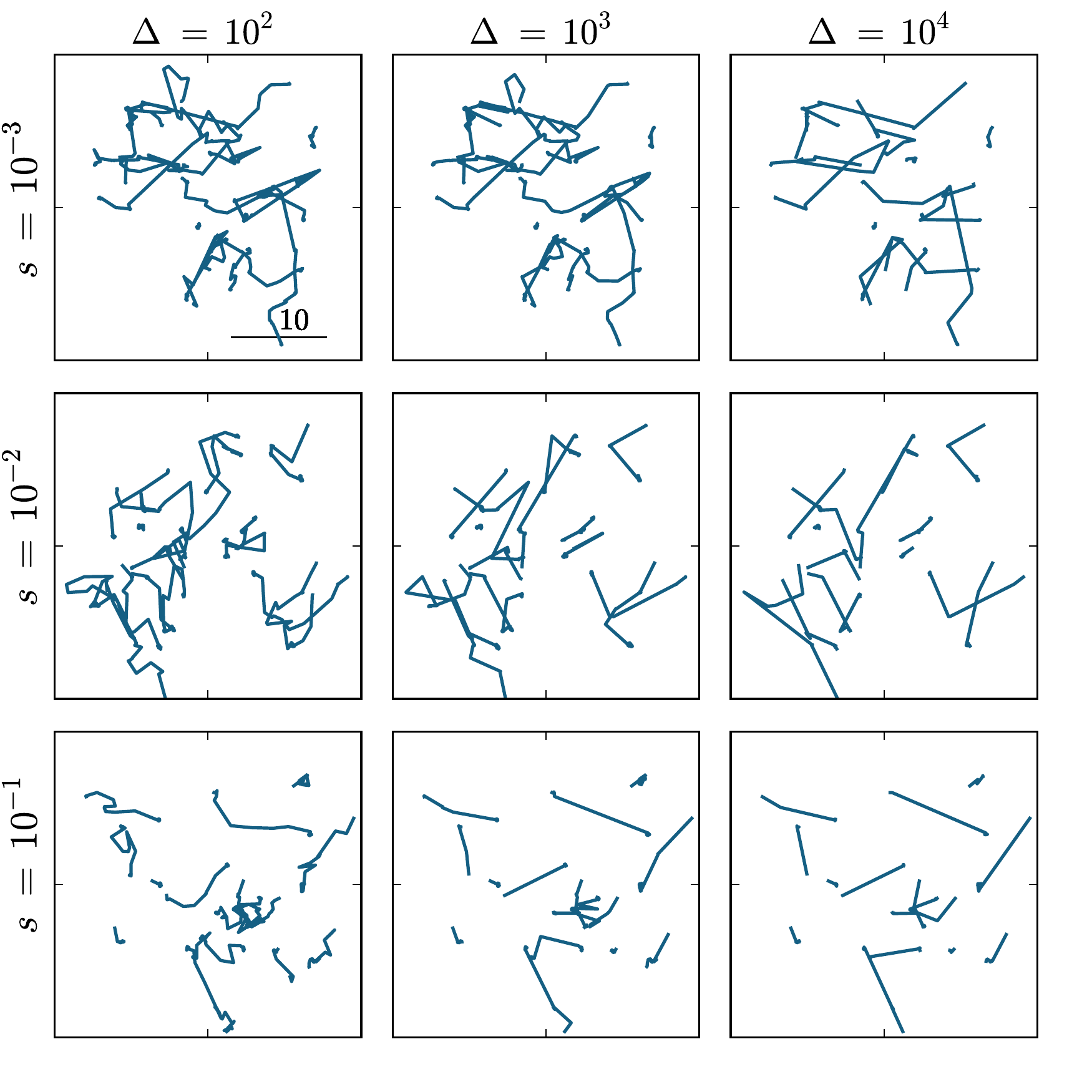}
 \end{center}
\caption{Example trajectories for binding radii $s = 0.001$, 0.01, and 0.1. The trajectories are coarse-grained by only showing points that are temporally separated by $\Delta$. 
The parameters used are the same as those in the main text (Table 1).}
\label{fig:s4}
\end{figure}

Additionally, we show the trajectories for different binding radii and temporal coarse-grainings $\Delta$ (Fig. \ref{fig:s4}). When $\Delta$ is  larger than the average trapping time in a vortex, the time-averaged MSD is diffusive. 

\begin{figure}[htb]
\begin{center}
 \includegraphics[width=0.75\textwidth]{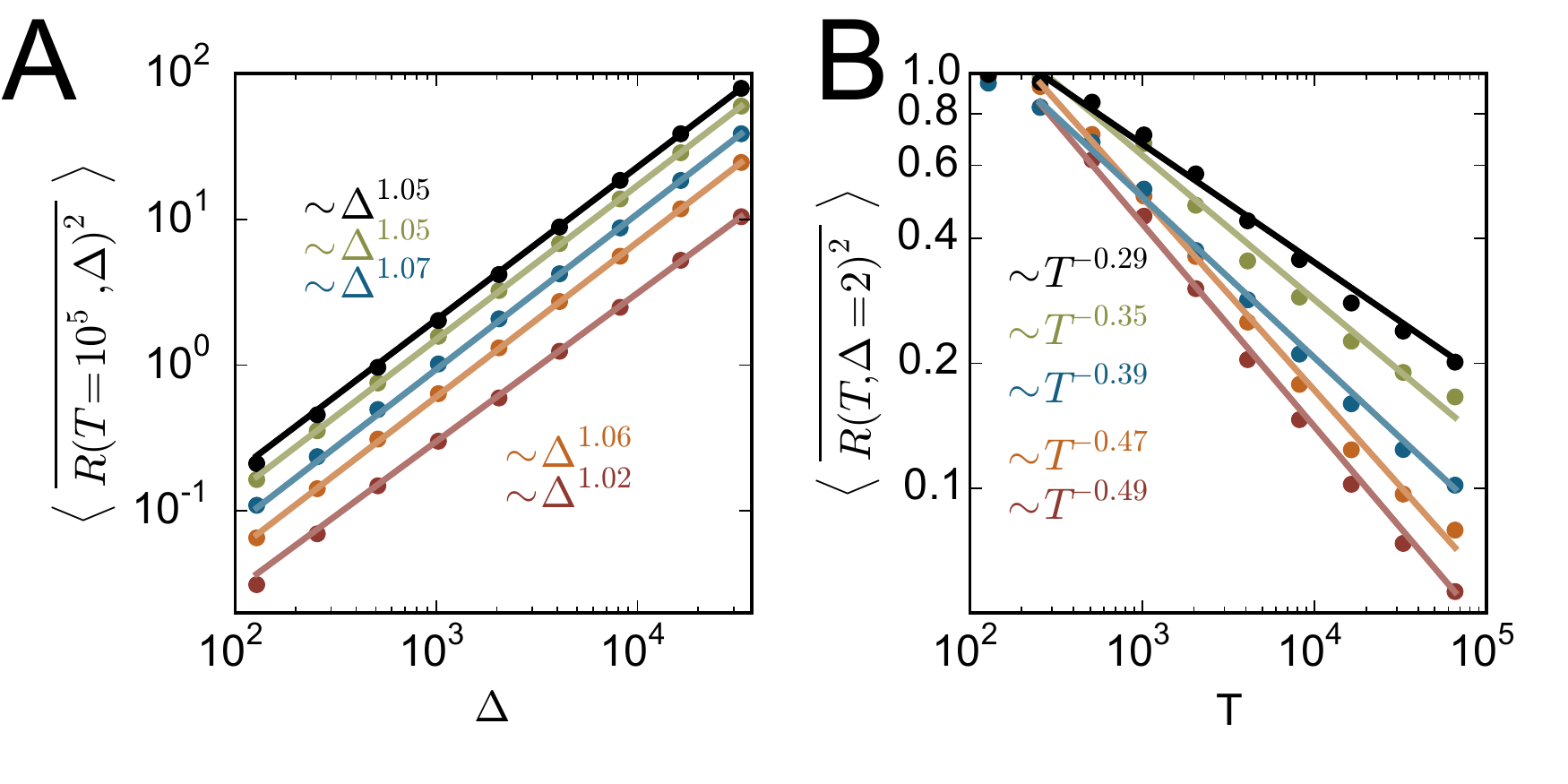}
 \end{center}
\caption{Mean-squared displacement for different filament network densities. (A) MSD as a function of lag time for network density (number of filaments per unit area) $d =1$ (red),  $d =2$  (orange),  $d =3$ (blue), and  $d =4$  (green) and  $d =5$  (black). (B) The MSD as a function of measurement time for the same densities as in (A).  Other parameters are the same as those in the main text (Table 1).}
\label{fig:s5}
\end{figure}

\section{Experimental details}
\begin{figure}[h]
\begin{center}
 \includegraphics[width=0.75\textwidth]{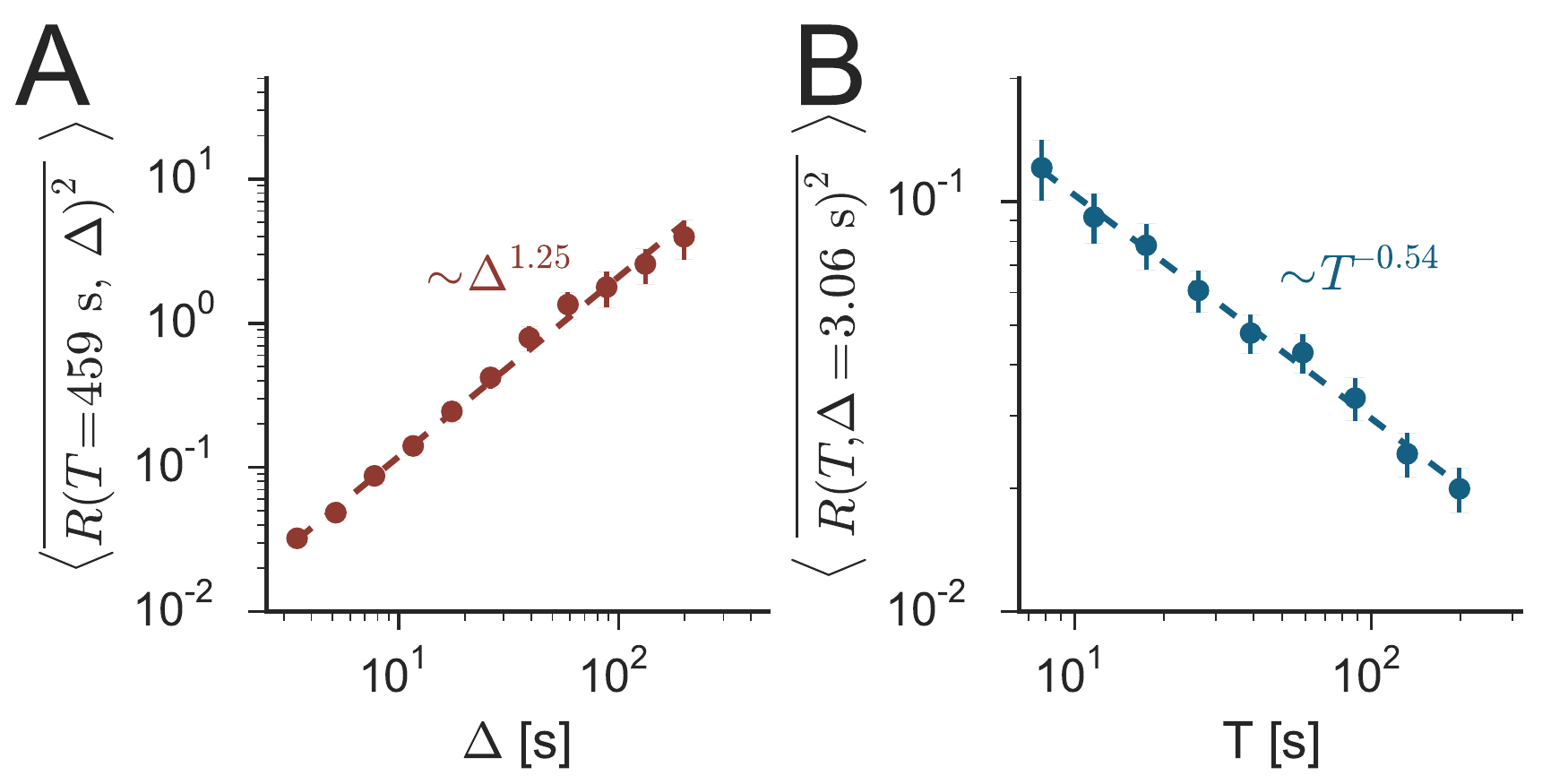}
 \end{center}
\caption{Mean-squared displacement for myosin II motors. (A) MSD as a function of lag time. (B) The MSD as a function of measurement time. }
\label{fig:exptmsd}
\end{figure}
Networks of bundled actin were polymerized in a thin layer at a surface of a flowcell. A supported lipid bilayer passivated the surface. The supported bilayer was formed by incubating a UV-ozone cleaned borosilicate coverslip (Fisherbrand) with 1 mM vesicle suspension. Vesicles were prepared by the standard method of extrusion (200 nm and 50 nm pore membranes, Liposofast extruder, Avestin) from dried films of phospholipid (1,2-dioleoyl-sn-glycero-3-phosphocholine, Avanti Polar Lipids) suspended in buffer (140 mM NaCl, 8.5 mM Na$_2$HPO$_4$, 1.5 mM NaH$_2$PO$_4$, pH 7.5). After a complete bilayer formed, excess vesicle suspension was exchanged with actin polymerization buffer (10 mM imidazole, 50 mM KCl, 0.2 mM EGTA pH 7.5, 300 $\mu$M ATP). Monomeric actin (rabbit skeletal muscle purified from acetone powder, Pel-freeze; 2.0 $\mu$M unlabeled and 0.64 $\mu$M labeled with the fluorophore tetramethylrhodamine-6-maleimide, Life Technologies) was added to initiate the polymerization of long, entangled actin. Depletion agent (0.3 wt \%, 15 centipoise methylcellulose, Sigma) crowded actin to the surface. We used an oxygen scavenging system (50 $\mu$M glucose, 0.5 vol \% $\beta$-mercaptoethanol, glucose oxidase, and catalase) to reduce photobleaching. After 30 min of polymerization, fimbrin (0.53 $\mu$M, pombe) was added to crosslink the actin filaments into a network of bundles. Thick filaments of myosin II were  polymerized in a similar manner by adding monomeric myosin II (20 nM, chicken skeletal muscle, fluorescently labeled with Alexa 647, Life Technologies) to a separate, actin-free solution. After 10 min of polymerization, myosin in ATP ($\sim$2\% of total sample volume) was gently mixed with the solution above the actin network, such that the final concentrations were 3.8 pM myosin and 2.3 mM ATP. Actin and myosin were imaged with a inverted microscope (Nikon Eclipse Ti-PFS) equipped with a spinning disk confocal head (CSUX, Yokogawa), 561 nm and 647 nm laser lines, 60$\times$/1.49 NA oil immersion objective (Zeiss), and a CCD camera (Coolsnap HQ2, Photometrics). Images, collected at 1.5 s intervals, began $\sim$10 min after myosin was added to the sample.

\begin{figure}[h]
\begin{center}
 \includegraphics[width=0.75\textwidth]{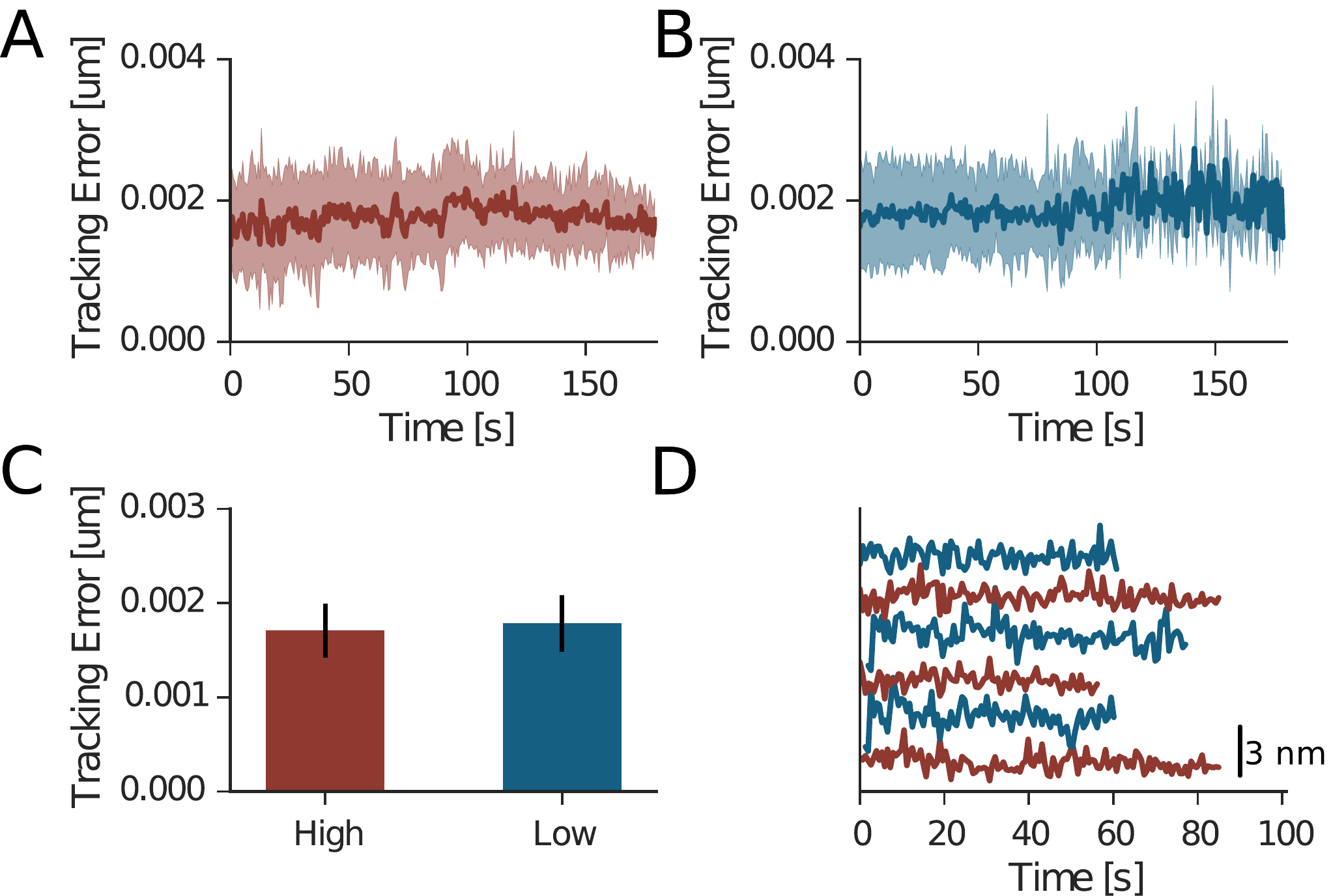}
 \end{center}
\caption{(A, B) Tracking accuracy for trajectories with high (red) or low (blue) fluorescence. The shaded area denotes the standard deviation. (C) Mean tracking accuracy for high and low fluorescence trajectories. Lines indicate standard deviations. (D) Tracking error for randomly selected trajectories over time. }
\label{fig:tracking}
\end{figure}

Single-particle trajectories were obtained using the Python-based implementation of the Crocker-Grier algorithm Trackpy. 
The settings are given in Tab.\ \ref{tab:trackpy_parameters}. The particle identification parameters were chosen such that all visible features were detected and the mask size was chosen large enough to obtain subpixel resolution (The histogram of the decimal values of the identified particles showed a flat distribution, indicating no bias within a pixel.). The linking values were set by looking at a previously manually tracked data set and using the largest displacement of that dataset as search range. All other parameters were set to the default values.  We quantify the static tracking error for brighter and dimmer objects (implemented in Trackpy). 
The Fig. \ref{fig:tracking} shows the ensemble averaged tracking error as a function of time.  There is no overall trend in the accuracy over the time of tracking in high (red) or low (blue) brightness fluorescent objects, and there is no statistical difference in the mean tracking error between the two populations (mean and standard deviation shown, unequal variances $t$-test $p$-value=0.0459)

\begin{table}
[tb]
%\begin{ruledtabular}
\begin{tabular}{lr} \hline
Parameter& Value \\ \hline
mask diameter & 19\\
minimum mass &650000\\
noise size &3\\
search range&15\\
memory&1
\end{tabular}
%\end{ruledtabular}
\caption{Particle tracking parameters for Trackpy.}
\label{tab:trackpy_parameters}

\end{table}

\begin{figure}[h]
\begin{center}
 \includegraphics[width=0.75\textwidth]{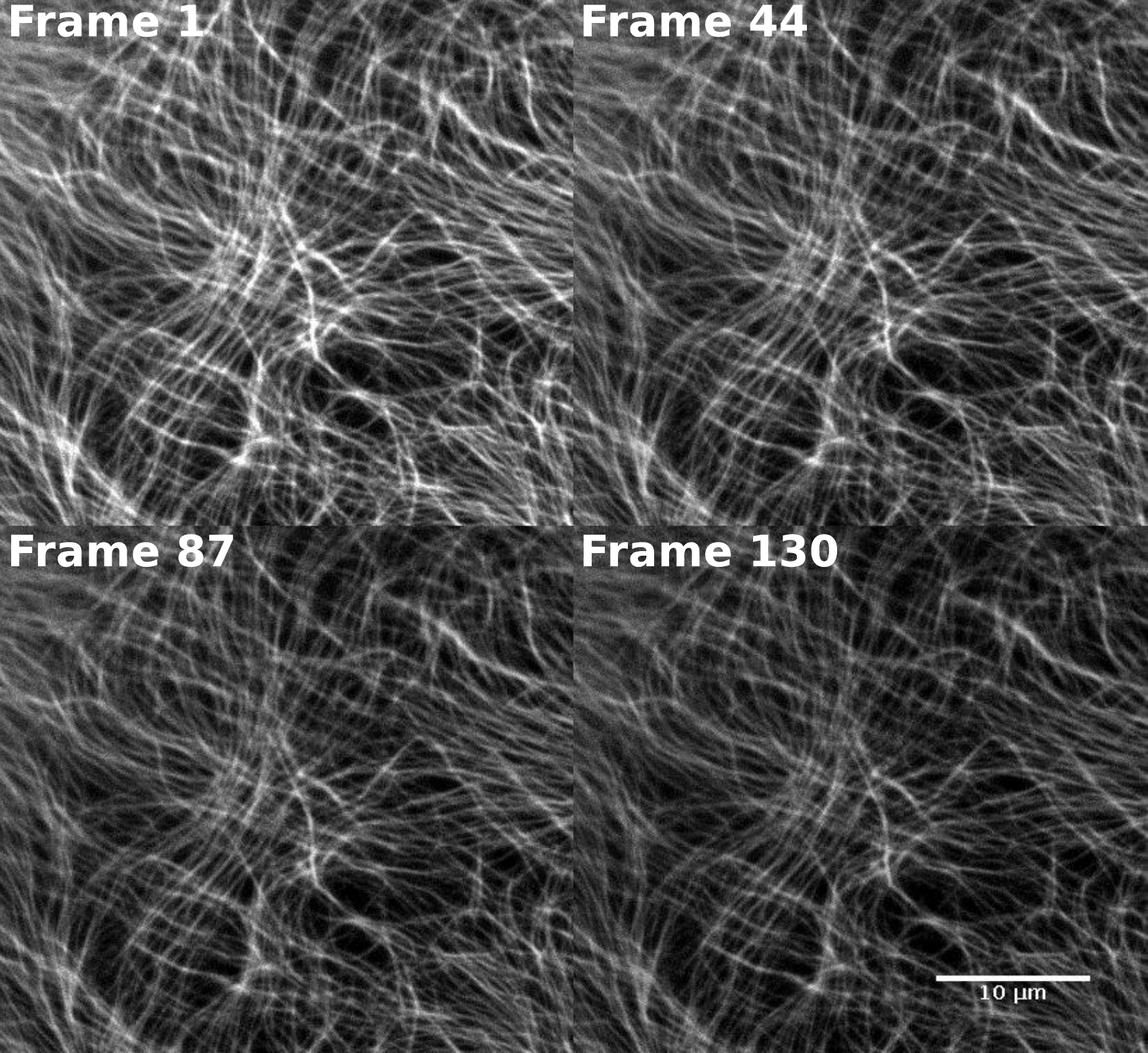}
 \end{center}
\caption{Fluorescently labelled actin network. The images show snapshots of the actin network while myosin II motors are moving on it (not shown). While the frames show some photobleaching over the course of the experiments, the network does not remodel.}
\label{fig:exptnetwork}
\end{figure}

\section{Analysis of trapped states}

Here, we consider the possibility that detachment could give rise to the power-law dwell times evidenced by the decay of the MSD with the measurement time.  If  detachment were responsible for the anomalous dynamics, the particles should undergo simple diffusion when apparently trapped.  To test for this possibility, we divided the trajectories between trapped and non-trapped periods and then analyzed their dynamics as follows.

\begin{figure}[hbt]
\includegraphics[width=0.75\textwidth]{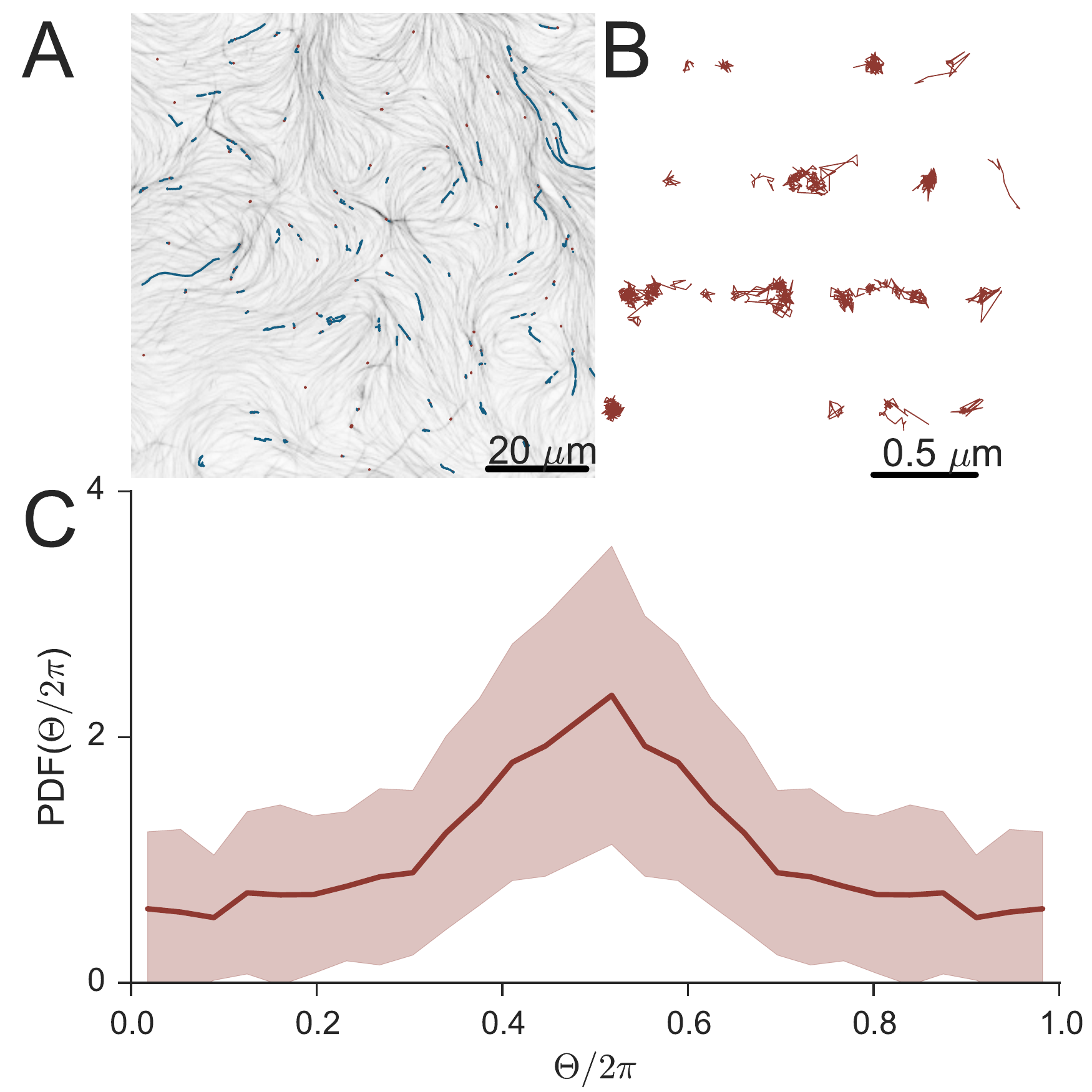}
\caption{The dynamics in apparent trapped states are not simple diffusion.  (A) Representative single-particle trajectories colored by whether they are assigned to be trapped (red) or not (blue) according to the criteria in the accompanying text.  The actin network is shown in gray in the background.  (B) Magnified view of selected trapped trajectory segments.  (C) Relative angle distribution of trapped trajectory segments for $\Delta = 1$ frame.  Shaded area indicates standard deviation.  \label{trappedfig}
}
\end{figure}

To define the trapped portions of single-particle trajectories, we calculated the average position of each particle (motor) every 40 frames. We then compared the instantaneous positions with the averaged ones. If a point in a trajectory deviated by less than 2.0 pixel widths (182 nm) from the corresponding average position, we count that frame as trapped. The results are shown in Figure \ref{trappedfig}.  Representative trapped trajectory segments are shown in red in the context of the non-trapped segments and the filament network (Figure \ref{trappedfig}A) and in a magnified view (Figure \ref{trappedfig}B). 

To test for simple diffusion with a time resolution of a few frames, we employ the relative angle distribution introduced by Burov \textit{et al.} \cite{burov2013distribution}.   The relative angle is defined by
%%%%%%%%%%%%%%%%%%%%%%%%%%%%%%%%%%%%%%%%%%%%
\begin{equation}
\cos\theta(t;\Delta)= \frac{\vec{v}(t;\Delta)\cdot  \vec{v}(t+\Delta;\Delta)}{| \vec{v}(t;\Delta)||   \vec{v}(t+\Delta;\Delta)|}.
\label{equ01}
\end{equation} 
%%%%%%%%%%%%%%%%%%%%%%%%%%%%%%%%%%%%%%%%%%%%%
$\vec{v}(t;\Delta)= \vec{x}(t+\Delta)- \vec{x}(t)$, $\vec{x}(t)$ is the position at time $t$, and $\Delta$ is the lag time as in the present paper.  Here we use $\Delta =1$ frame.  We build a histogram of these values for the trapped (red) trajectory segments and normalize such that it integrates to one.

The relative angle distribution for the trapped trajectory segments is shown in  Figure \ref{trappedfig}C.  If the motion were simple diffusion, the distribution would be flat, because there would be no directional correlation between steps of the random walk.  Instead, we see that it is peaked at $\Theta = \pi$, which indicates that the particles are making reversals.  The only way that this could happen from detachment would be if the particles were caged when off the filaments.  However, this cannot be the case, as the prevalent switching between trapped and non-trapped states shows that the majority of the particles are unimpeded by obstacles.  The only possibility consistent with  Figure \ref{trappedfig}C is that the particles are switching direction while on the filaments, owing to the motor dynamics---this is the cycling mechanism that we propose.

\end{document}